\documentclass[lettersize,journal,a4paper]{IEEEtran}
\usepackage{amsmath}
\usepackage{amsthm}
 
\theoremstyle{plain}

\newtheorem{remark}{Remark}[section]

\newtheorem{lemma}{Lemma}[section]

\newtheorem{theorem}{Theorem}[section]

\newtheorem{assumption}{Assumption}

\newtheorem{corollary}{Corollary}[theorem]

\usepackage{color}
\usepackage{graphicx} 
\usepackage{bm}
\usepackage{multicol}
\usepackage{setspace}
\usepackage{subfigure}
\usepackage{epstopdf}
\usepackage{fancyhdr} 
\usepackage{indentfirst}
\usepackage{enumerate}
\usepackage{amssymb}
\usepackage{stfloats}
\usepackage{bm}
\usepackage{threeparttable}
\usepackage{CJK}
\usepackage[justification=centering]{caption}
\usepackage{makecell}
\usepackage{caption}
\usepackage{cases}
\usepackage{algorithm}
\usepackage{algpseudocode}
\usepackage{graphics}
\usepackage{subfig}
\usepackage{epsfig}
\usepackage{cite}

\begin{document}
	\title{{D\(^2\)}-JSCC:  Digital Deep  Joint Source-channel Coding for Semantic Communications}
\author{
	{Jianhao~Huang, Kai~Yuan, Chuan~Huang, ~\IEEEmembership{Member,~IEEE}, and Kaibin Huang, ~\IEEEmembership{Fellow,~IEEE} }
	\thanks{
		
				{K. Yuan and C. Huang  	are with the School of Science and Engineering and the Future Network of Intelligence Institute, the Chinese University of Hong Kong, Shenzhen, 518172 China. Emails: kaiyuan3@link.cuhk.edu.cn and huangchuan@cuhk.edu.cn.  }
				
		{J. Huang and K. Huang are with the Department of Electrical and Electronic Engineering, The University of Hong Kong, Hong Kong. Emails: jianhaoh@hku.hk  and huangkb@eee.hku.hk (\emph{Corresponding author: Kaibin Huang}). }

	}
}
	
	\maketitle
	\thispagestyle{empty}
	\begin{abstract}
Semantic communications (SemCom) have emerged as a new paradigm for supporting \emph{sixth-generation} (6G) applications, where semantic features of raw data are extracted and transmitted  using artificial intelligence (AI) algorithms  to attain high communication efficiencies. Most existing SemCom techniques rely on deep neural networks (DNNs) to implement analog (or semi-analog) source-channel mappings. 
These operations, however, are not compatible with existing digital communication architectures. 
To address this issue, we propose in this paper a novel framework of digital deep joint source-channel coding (D$^2$-JSCC) targeting image transmission in SemCom. 
The framework features digital source and channel codings that are jointly optimized to reduce the end-to-end (E2E) distortion. 
First, deep source coding with an adaptive density model is designed to efficiently extract and encode semantic features according to their different distributions. 
Second,  channel  block coding is employed to protect encoded features against channel distortion. 
To facilitate their joint design, the E2E distortion is characterized as a function of the source and channel rates via the analysis of the Bayesian model of the D$^2$-JSCC system and validated Lipschitz assumption on the DNNs. 
Then to minimize the E2E distortion, we propose an efficient two-step algorithm to find the optimal trade-off between the source and channel rates for a given channel signal-to-noise ratio (SNR). 
In the first step, the source encoder is initially optimized by selecting a DNN model with a suitable source rate from a designed look-up table consisting of a set of trained models. 
In the second step, the preceding DNNs (source encoder) are retrained to adapt to the channel SNR so as to achieve the optimal E2E performance. 
Via experiments on simulating the D$^2$-JSCC with different channel codes and real datasets, the proposed framework is observed to outperform the classic deep JSCC. Furthermore, due to the source-channel integrated design, D$^2$-JSCC is found to be free from the undesirable \emph{cliff effect} and \emph{leveling-off effect}, which commonly exist for digital systems designed based on the separation approach. 

	\end{abstract}
	

	\section{Introduction}
	The sixth-generation (6G) mobile networks are being developed to embrace  the explosive growth of the population of edge devices and support a broad range of emerging applications, such as autonomous vehicles, surveillance, and virtual/augmented reality (VR/AR) \cite{saad2019vision,zhu2020toward,lin2024efficient}. This poses tremendous challenges of utilizing limited spectrum resources to meet much more stringent performance requirements than those for fifth-generation (5G) \cite{letaief2019roadmap,lin2023split}. As a promising solution empowered by  artificial intelligence (AI), semantic communications (SemCom) start a new paradigm to effectively extract and transmit  semantic features of raw data, thereby substantially reducing the communication overhead \cite{zhang2022toward,gunduz2022beyond,10158994,10287247}. Unlike the  Shannon’s separation approach,  SemCom integrates the source and channel coding for boosting the end-to-end (E2E) system capabilities \cite{qin2021semantic,gunduz2022beyond}.  However, the intricate nature of  channel environment and the constraints imposed by existing digital hardware present  challenges in the development   of AI-empowered transceivers for SemCom.
	
	One representative technology for SemCom  involves the use of AI to enhance the integrated design of source and channel codes, named as joint source-channel coding (JSCC), which is a classical topic in  information  and coding theories. Generally, the traditional JSCC schemes can be divided into two categories: analog JSCC \cite{ramstad2002shannon,10189867} and digital JSCC \cite{fresia2010joint,quantization,nosratinia2003source,hamzaoui2005optimized}. In the former, the continuous source symbols are directly mapped to the analog signals for transmission by a linear/nonlinear function, e.g., Shannon-Kotel’nikov mapping \cite{10189867}. Despite its capability of achieving the rate-distortion bounds, the analog JSCC is hard to implement in practice. Digital JSCC aims to be compatible with the existing digital communication systems. Examples of this approach include: 1) optimal codeword assignment for source data \cite{fresia2010joint}; 2) optimal quantizer design for noisy channel \cite{quantization}; 3) joint source-channel rate control \cite{nosratinia2003source}; 4) unequal error protection \cite{hamzaoui2005optimized}. However, these traditional JSCC schemes are generally based an oversimplified probabilistic model of source data without considering their semantic aspects.  Addressing this limitation using AI ushers in a new era of JSCC. 	
	
	Recently, the impressive capability of deep learning methods for nonlinear mappings has sparked significant interests in implementing analog JSCC for real-world data. By using the deep neural networks (DNNs), a so called deep JSCC scheme directly maps the source data (e.g., image \cite{dai2022nonlinear,kurka2019}, text \cite{9398576}), into a reduced-dimensional feature space for transmissions over analog channels.  The early work  on  deep JSCC schemes has  shown a higher compression efficiency than those of the classical separation-based JPEG/JPEG2000/BPG compression schemes combined with practical channel codes \cite{dai2022nonlinear,kurka2019}. However, the analog nature of the schemes  make it incompatible with modern communication hardware that is prevalently digital. This motivates researchers to transform  continuous-valued outputs from DNNs into discrete constellation symbols for transmission, thereby establishing a \emph{semi-analog} deep JSCC framework \cite{9998051,bo2023joint}. In relevant schemes, an additional DNN-based modulation block is introduced to generate transition probabilities from learned features to the constellation symbols. The DNNs for encoder, modulation, and decoder are jointly trained to optimize the E2E performance. However, the semi-analog JSCC still lacks the digital channel encoder/decoder and thus is incompatible with the standard systems. Furthermore, the DNN training  relies on  off-line back propagation over a certain number of channel samples, and thus is sensitive to the variation of channel statistics and machine task. For mission-critical applications, it is usually infeasible to collect new channel samples and retrain the models from time to time. 
	
	By involving the DNNs into the digital JSCC, several research works make efforts to combine the deep-learning based source coding with  digital channel coding for SemCom \cite{he2023rate,liu2024ofdm}. The key idea is to utilize a fixed  DNN to extract  data features and  analyze their semantic importance, which enables the channel encoder to identify the critical part of features for unequal error protection. Specifically, the authors in \cite{he2023rate}  proposed a rate-adaptive coding mechanism to unequally assign channel  rates for different mutil-modal data according to their semantic importance.    The authors in \cite{liu2024ofdm} proposed a deep reinforcement learning based algorithm for  sub-carrier and bit allocation by characterizing the correlation importance among  semantic features and tasks.
However, these works assume that the deep source coding is independent of  the channel statistics. Consequently, the performances of these schemes are limited by the pre-trained DNNs. 
	
	In this paper, we aim to propose a novel framework of  \emph{digital deep joint source-channel coding} (D$^2$-JSCC) to address the image transmission problem in SemCom.  In particular, we consider a  point-to-point SemCom system,  where the transmitter utilizes the DNNs to extract the low-dimensional features of  image data  and sends them to the receiver for recovery.  Different from traditional deep JSCC schemes, D$^2$-JSCC utilizes deep source coding to encode semantic features, combined with  digital channel coding to protect the coded bits from channel errors. It then facilitates their joint optimizations to minimize the E2E distortion. 
The D$^2$-JSCC  addresses the following open problems in the digital SemCom: 
1) The quantization and digital channel encoding/decoding are discrete functions, which makes it difficult to optimize the DNNs by gradient descent algorithm \cite{lecun2015deep} in an E2E manner;
2) The intractability of DNNs stymies the derivations of  a closed-form expression of E2E distortion,  which is essential for the optimization of  channel coding.

Specifically, the key contributions and findings of this paper are summarized as follows:
\begin{itemize}
	\item \textbf{D$^2$-JSCC Architecture}:   We propose a novel D$^2$-JSCC architecture for SemCom, which combines the deep source coding with the digital channel coding. First,  the deep source coding with an adaptive density model \cite{Balle2018}  is designed to efficiently extract and encode semantic features of data. The adaptive density model learns the probability density function (PDF) of the features as \emph{side information}, which helps  to encode them with a higher coding efficiency. Then, digital channel block coding is employed to safeguard the encoded features for transmissions. Based on the architecture, an E2E distortion minimization problem is formulated. 
	\item \textbf{E2E Distortion Approximation}: To characterize the E2E distortion, we propose a Bayesian approximation of  the feature space and make a Lipschitz assumption on  DNNs. Based on these, the intractable E2E distortion can be approximately derived as a function with respect to (w.r.t.) the parameters of DNNs and channel rate. From the observation of the E2E distortion, it is found  that the key problem of minimizing the E2E distortion is to jointly adapt the source-channel rates  to the channel signal-to-noise ratio (SNR).
	\item \textbf{Optimal Rate Control}: To minimize the E2E distortion, we propose an efficient two-step algorithm to balance the trade-off between the source and channel rates for a  given channel SNR.   In the first step, we derive the optimal channel rate for a given source rate and then optimize  the deep source coders by selecting a DNN model with a suitable source rate from a pre-designed look-up table. 
	  In the second step, the selected DNN model is retrained to adapt to the channel SNR, thereby achieving  the close-to-optimal E2E performance. 
	 It is worth mentioning  that the training of the deep source coders depends not on  channel samples but on channel statistical information, i.e., SNR. This makes this algorithm practical  even for time-sensitive systems.
	\item \textbf{Experiments}: Experimental results reveal that the proposed D$^2$-JSCC mitigates the “leveling-off effect” and “cliff effect” commonly existing  for  digital system\footnote{The ``cliff effect'' occurs when the channel SNR falls beneath a certain threshold and the E2E performance degrades drastically.  The ``leveling-off effect'' refers to the fact that the E2E performance remains constant even when the channel SNR is increased  above the threshold.}, since the  scheme can adaptively optimize the  source and channel coding according to different channel SNRs. In addition, 
the proposed scheme outperforms classic deep JSCC scheme. The reason for this is that the latter fixes the number of transmitted symbols for all images, while the former with an adaptive model has the capability to vary the number of symbols based on the image content and channel SNR.   Furthermore, we observe that 
 as the block length increases, the E2E performance of D$^2$-JSCC   increases and approaches  that of the separate source-channel coding with capacity achieving code. It implies that the channel coding length benefits the D$^2$-JSCC, while this phenomenon does not occur in the deep JSCC.
\end{itemize}

The remainder of this paper is organized as follows. The architecture of the D$^2$-JSCC and the problem of E2E distortion minimization are introduced in Section II. The  E2E distortion is characterized in Section III and the algorithm for the said problem is proposed in Section IV.    Experimental results  are presented in Section V, followed by concluding remarks in Section VI. 

Notations:  We utilize lowercase and uppercase letters, e.g., $x$ and $M$, to denote  scalars, and use boldface lowercase letters, e.g., $\bm{x}$, to denote vectors. $\mathbb{Z}$,  $\mathbb{R}$, and  $\mathbb{C}$, denote the sets of all integer, real, and complex values, respectively.
$||\bm{x}||$ denotes the $2$-norm  of vector $\bm{x}$. $\bm{x}^T$ and $\bm{x}^H$ denote the  transpose and conjugate transpose of vector $\bm{x}$, respectively.  $p_{\bm{x}}(\bm{x})$ denotes the PDF of the continuous random variable $\bm{x}$.  $P_{\bm{y}}(\bm{y})$ denotes the probability mass function (PMF) of the discrete random variable $\bm{y}$.   $O(\cdot)$ denotes the big O  notation.  $\text{log}(\cdot)$ and $\text{log}_2(\cdot)$ are the logarithm functions with base $e$ and $2$, respectively. $\text{Tr}(\bm{X})$ and $\text{det}(\bm{X})$ denote the trace and determinant of matrix $\bm{X}$, respectively.  

 \begin{figure*}[t]
	\normalsize
	\setlength{\abovecaptionskip}{+0.3cm}
	\setlength{\belowcaptionskip}{-0.1cm}
	\centering
		\includegraphics[width=6.6in]{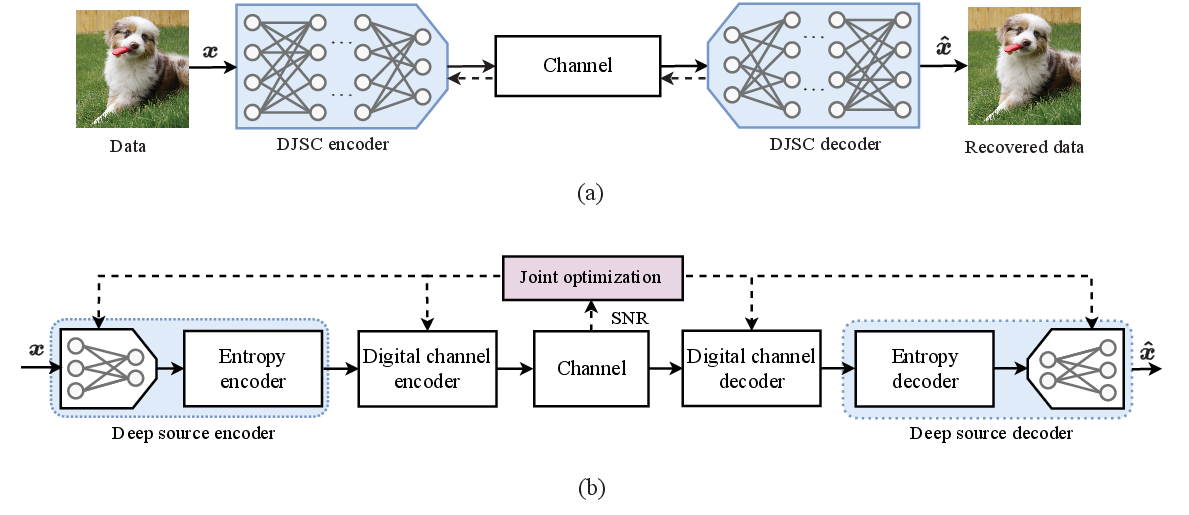}
		\captionsetup{justification=justified}
\caption{Architecture comparison of the JSCC schemes empowered by deep learning techniques: (a) traditional deep JSCC; (b) proposed D$^2$-JSCC. The solid and dashed arrows represent the directions of signal flows and optimization paths, respectively.  }
		\label{architecture}
\end{figure*}

 \begin{figure*}[t]
	\normalsize
	\setlength{\abovecaptionskip}{+0.3cm}
	\setlength{\belowcaptionskip}{-0.1cm}
	\centering
		\includegraphics[width=6.4in]{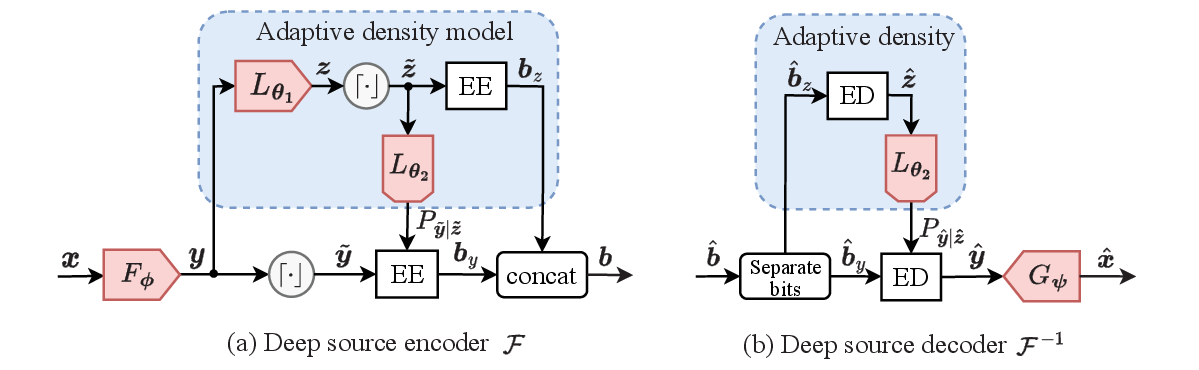}
\captionsetup{justification=justified}
\caption{Architectures of the deep source encoder and decoder using adaptive density model. $\left\lceil \cdot \right\rfloor$, EE, and ED represent the quantization, entropy encoder, and entropy decoder, respectively. }
\label{source architecture}
\end{figure*}

	\section{D$^2$-JSCC Architecture and Problem Formulation}
	
	Consider a point-to-point SemCom system where the transmitter aims to compress a $M$-dimensional   image vector $\bm{x}$, and send it to the receiver for recovery. Due to the bandwidth limitation, $\bm{x}$ needs to be compressed and encoded into digital symbols for transmission.  To boost the E2E performance of the SemCom system, a novel  D$^2$-JSCC framework, which integrates the digital communication architecture with deep learning techniques, is proposed as shown in Fig. \ref{architecture}(b).  For comparison, the traditional deep JSCC architecture is shown in Fig. \ref{architecture}(a).  The proposed D$^2$-JSCC framework combines  digital source and channel coding, which are described in separate subsections. In the last two subsections, we introduce the overall transmission process of the considered SemCom system and formulate the optimization problem. 

	\subsection{Deep Source Coding}

 Source coding aims to  compress data into bit streams within a certain amount of distortion,  consisting of an encoding function 
	$\mathcal{F}:  \mathcal{X}^{M} \rightarrow \{0,1\}^{B}$,
and a decoding function $
	\mathcal{F}^{-1}:  \{0,1\}^{B} \rightarrow \mathcal{X}^{M}$,
where $\mathcal{X}^{M}$ is the set of all possible data $\bm{x}$ of dimension $M$ and $B$ is the number of  coded bits. The distortion of source coding, denoted by $\mathcal{D}_s$, is measured using average mean square error (MSE) metric, i.e.,
\begin{align}\label{source-d}
	\mathcal{D}_s=\mathbb{E}_{\bm{x}}\left\{\frac{1}{M}||\bm{x}-\mathcal{F}^{-1}(\mathcal{F}(\bm{x}))||^2\right\}.
\end{align} 

Since the data's distribution is usually intractable, we employ the method of deep source coding  \cite{Balle2017,Balle2018,9242247,10175391} that utilizes DNNs to learn the close-to-optimal encoding and decoding functions according to a certain number of data samples. Specifically, this method leverages DNNs to extract semantic features from  data and integrates density models to adaptively encode them, ultimately approaching the optimal coding performance \cite{yang2022towards,li2023fundamental}.  In the following, we introduce  the deep source encoder and decoder, as illustrated in Fig. \ref{source architecture}.

\begin{itemize}
	\item \textbf{Deep source encoder}: First, the input image sample, $\bm{x} \in \mathcal{X}^{M}$, with element value ranging  from $0$ to $1$, is mapped to a $K$-dimensional continuous vector, $\bm{y}$, by the feature extraction function, $F_{\bm{\phi}}$, parameterized by the variable $\phi$. For lossy compression, $K$ is much smaller than $M$.  Then,  $\bm{y}$ is quantized as  a discrete vector, $\tilde{\bm{y}}\in \mathbb{Z}^{K}$, i.e., $\tilde{\bm{y}}=\left\lceil \bm{y} \right\rfloor$, where $\left\lceil \cdot \right\rfloor$ denotes the uniform scalar quantization with step size being one \cite{Balle2017, Balle2018}. Next,  lossless entropy encoding (e.g., arithmetic encoding \cite{witten1987arithmetic}), is employed to encode the quantized vector  $\tilde{\bm{y}}$ into a bit stream, $\bm{b}_y \in \{0,1\}^{B_y}$, according to its PMF, $P_{\tilde{\bm{y}}}(\tilde{\bm{y}})$, with the bit length $B_y\approx -\log_2 P_{\tilde{\bm{y}}}(\tilde{\bm{y}})$.
 It is worth mentioning that for image data, the PMF $P_{\tilde{\bm{y}}}(\tilde{\bm{y}})$ captures the spatial dependencies of the feature vector $\bm{y}$. Hence, accurately learning $P_{\tilde{\bm{y}}}(\tilde{\bm{y}})$ can significantly reduce the redundancy of feature representation, leading to a lower source rate. A standard way to model the dependencies among the feature elements is to introduce latent variables conditioned on which the elements are assumed to be independent \cite{Balle2018}. To this end, we introduce an additional set of random variables, $\bm{\tilde{z}}=[\tilde{z}_1,\tilde{z}_2,\cdots,\tilde{z}_{D}]^T\in \mathbb{Z}^{D}$ with $D<K$, to capture the spatial dependencies of $\tilde{\bm{y}}$. Then, the PMF $P_{\tilde{\bm{y}}}(\tilde{\bm{y}})$ used for encoding feature $\tilde{\bm{y}}$  is replaced with $P_{\tilde{\bm{y}}|\tilde{\bm{z}}}(\tilde{\bm{y}}|\tilde{\bm{z}})$. The latent variable $\bm{\tilde{z}}$ is  often referred to as the \emph{side information} of features, and needs to be transmitted to receiver. 
 
 Here, we introduce the \emph{adaptive density model} that extracts  $\bm{\tilde{z}}$ from $\bm{y}$ and calculates $P_{\tilde{\bm{y}}|\tilde{\bm{z}}}(\tilde{\bm{y}}|\tilde{\bm{z}})$. In the  model,  $\bm{y}$ is fed into a function $L_{\theta_1}(\bm{y})$ with parameter $\theta_1$ to extract the continuous vector, $\bm{z} \in \mathbb{R}^{D}$, which is quantized as  $\tilde{\bm{z}}\in \mathbb{Z}^{D}$, i.e., $\tilde{\bm{z}}=\left\lceil \bm{z} \right\rfloor$. The features, $\{y_i\}$, conditioned on  $\bm{\tilde{z}}$ can be modeled as the independent while not identically distributed  (i.n.i.d.) Gaussian random variable  with mean $u_i$ and  variance $\sigma_{i}^2$. In other words, 
 \begin{align} \label{norm:1}
 	p_{y_i|\bm{z}}(y_i|\bm{\tilde{z}})=\mathcal{N}(y_i;u_i,\sigma_i^2),
 \end{align}
 where $\mathcal{N}(a;u,\sigma)$ denotes the  PDF of a Gaussian distribution with mean $u$ and variance $\sigma$, evaluated at the point $a$.  $u_i$ and $\sigma_{i}$ are estimated by applying a transform function, $L_{\theta_2}$, with a parameter $\theta_2$ to $\bm{\tilde{z}}$, i.e., $[\bm{u},\bm{\sigma}]=L_{\theta_2}(\bm{\tilde{z}})$ with $\bm{u}=[u_1,u_2,\cdots,u_K]^T$ and $\bm{\sigma}=[\sigma_1,\sigma_2,\cdots,\sigma_{K}]^T$. The conditional i.n.i.d. distribution of $\tilde{\bm{y}}$ can be expressed as 
  \begin{align} \label{den:y}
 		P_{\bm{\tilde{y}}|\bm{\tilde{z}}}(\bm{\tilde{y}}=\bm{k}|\bm{\tilde{z}})= \prod_{i=1}^{K} \left\{\int_{k_i-0.5}^{k_i+0.5}\mathcal{N}(y_i;u_i,\sigma_i^2)dy_i\right\}, 
 \end{align}
with $\bm{k}=[k_1,k_2,\cdots,k_K]^T\in \mathbb{Z}^{K}$.	The side information, $\tilde{\bm{z}}$, is encoded into bits $\bm{b}_z \in \{0,1\}^{B_z}$ by the entropy encoder according to its PMF, $P_{\tilde{\bm{z}}}(\tilde{\bm{z}})$, which is  computed using  the non-parametric fully-factorized density model \cite{Balle2017}. Finally, the bits of features and side information are concatenated into the bit streams, $\bm{b} \in\{0,1\}^{B}$,  with $B=B_y+B_z$ for transmission. The expected source rate of encoding data, $\bm{x}$,  is defined as the entropy of $(\tilde{\bm{y}},\tilde{\bm{z}})$, i.e., $R_s=\mathbb{E}_{\bm{x}}\{B\}=H(\tilde{\bm{y}},\tilde{\bm{z}})=\mathbb{E}\left\{-\log_2 P_{\tilde{\bm{y}}|\bm{\tilde{z}}}(\tilde{\bm{y}}|\bm{\tilde{z}})-\log_2 P_{\tilde{\bm{z}}}(\tilde{\bm{z}})\right\}$ \cite{Balle2018,9242247}.   In conclusion, the encoding function $\mathcal{F}$ can be specified in terms of parameters $\{\bm{\phi},\theta_1,\theta_2\}$, i.e., $
	\bm{b}=\mathcal{F}(\bm{x};\bm{\Phi})$,
with $\bm{\Phi}=\{\phi,\theta_1,\theta_2\}$.

\item \textbf{Deep source decoder}: As shown in Fig. \ref{source architecture}(b), the received bits $\hat{\bm{b}}$ are separated into two parts: feature bits $\hat{\bm{b}}_y$ and side information bits $\hat{\bm{b}}_z$.  First,  $\hat{\bm{b}}_z$ is fed into the entropy decoder to decode the side information, $\hat{\bm{z}}$, according to the shared PMF $P_{\tilde{\bm{z}}}(\tilde{\bm{z}})$. Then, $\hat{\bm{z}}$ is fed into the function $L_{\theta_2}$ to compute the PMF, $P_{\hat{\bm{y}}|\hat{\bm{z}}}(\hat{\bm{y}}|\hat{\bm{z}})$, given in \eqref{den:y}. With $P_{\hat{\bm{y}}|\hat{\bm{z}}}(\hat{\bm{y}}|\hat{\bm{z}})$ and  $\hat{\bm{b}}_y$,  the feature vector, $\hat{\bm{y}}$, is decoded by utilizing the entropy decoder. Finally, the decoded feature vector, $\hat{\bm{y}}$, is input into the recovery function  $G_{\psi}$ parameterized by the variable $\psi$ to recover the image data $\hat{\bm{x}}$: $\hat{\bm{x}}=G_{\psi}(\hat{\bm{y}})$. Function   $G_{\psi}$ is designed to be the inverse function of $F_{\phi}$.  In a nutshell, the source decoding function, $\mathcal{F}^{-1}$, can be specified  in terms of parameters $\{\theta_2, \psi\}$ as 
$
	\hat{\bm{x}}=\mathcal{F}^{-1}(\hat{\bm{b}};\theta_2,\psi)
$ with $\mathcal{F}$ being the encoding function. 
\end{itemize}

In the  deep source encoder and decoder, the parameterized functions $\{F_{\phi},L_{\theta_1},L_{\theta_2},G_{\phi}\}$ are designed by using DNNs, whose architecture and training process will be introduced in the sequel sections.
\addtolength{\topmargin}{0.02in}	
\subsection{Digital Channel Coding}
The purpose of the digital channel coding is to protect the data bits delivered from source coding against channel errors. 
Without loss of generality, we consider an arbitrary  $(N,L)$ block code (e.g., polar code with binary phase shift keying (BPSK) modulation),  consisting of a channel encoder,
$
\mathcal{C}: \{0,1\}^{N} \rightarrow \mathcal{S}^{L},	
$
and a channel decoder
$
\mathcal{C}^{-1}: 	\mathcal{R}^{L} \rightarrow \{0,1\}^{N},
$
where $\mathcal{S}^{L} \subset \mathbb{C}^{L}$,  $\mathcal{R}^{L} \subset \mathbb{C}^{L}$, $L$, and $N$ denote the codebook of  transmitted symbols,  the set of received symbols, the block length, and the length of data bits, respectively. The channel rate, $R_c$, for the block code is calculated as $R_c=\frac{N}{L}$ and needs to be designed. Assuming that the transmitted symbols are equiprobable, the block error probability of the code  can be characterized as a function of the channel rate. For example, in the Additive White Gaussian Noise (AWGN) channel with an SNR of $\gamma$, the average block error probability with random coding and maximum likelihood (ML) decoder  is approximated by \cite{polyanskiy2010channel}
\begin{align}\label{randomc}
	\rho=Q\left(\frac{\sqrt{L}\left(\log_2(1+\gamma)-R_c\right)}{\sqrt{\left(1-\frac{1}{(1+\gamma)^2}\right)\log_2^2(e)}}\right), \text{for  large}\ L.
\end{align}
For  practical channel coding and modulation, the average block error probability  can be approximated as  \cite{nosratinia2003source,arikan2009channel}
\begin{align}\label{polarc}
	\rho=e^{\beta_{1}R_c+\beta_2},
\end{align}
where the parameters, $\beta_{1} >0$, and, $\beta_2 \in \mathbb{R}$, which depend on the SNR $\gamma$, channel code type, and block length $L$, can be easily estimated by offline simulations \cite{nosratinia2003source}.

\subsection{Transmission Process}
	Based on the preceding coding schemes, the overall transmission process  of the considered SemCom system is described as follows. At the transmitter side, the input data vector, $\bm{x}$, is encoded into the bit stream $\bm{b}\in \{0,1\}^{B}$ by the deep source encoder, i.e., $\bm{b}=\mathcal{F}(\bm{x};\bm{\Phi})$. Then,  $\bm{b}$ is divided into multiple packets of equal  length of  $N$ bits for transmission. For each packet, the common $(N,L)$ block code with the rate $R_c$ is employed to encode  the data bits into a symbol sequence  of length $L$, represented by $\bm{s}_i \in \mathcal{S}^{L}$ for packet $i$. The transmitted symbols satisfy the unit power constraint, i.e., $\frac{1}{L}\mathbb{E}(\bm{s}_i^H\bm{s}_i)=1$. The total number of transmitted packets is calculated by $T=\lceil\frac{B}{LR_c}\rceil$, where $\lceil\cdot\rceil$ denotes the rounding-up operation.   The bandwidth ratio is defined as   $\frac{TL}{M}$ to measure the average channel uses for transmitting one element of  $\bm{x}$.

The input-and-output relationship of the point-to-point channel can be expressed as 
\begin{align}\label{signal}
	r_{i,j}=h_i \sqrt{p_i} s_{i,j}+n_{i,j},\ i =1,2,\cdots,T,\  j =1,2,\cdots,L,
\end{align}
where ${\bm{s}}_i=[{s}_{i,1},{s}_{i,2},\cdots,{s}_{i,L}]^T$ denotes the  symbols in packet $i$, $\bm{r}_i=[r_{i,1},r_{i,2},\cdots,r_{i,L}]^T$ denotes the received symbols,  $p_i$  denotes the transmission power, and $\bm{n}_i=[n_{i,1},n_{i,2},\cdots,n_{i,L}]^T$ is the independent and identically distributed (i.i.d.) 
circularly symmetric complex Gaussian (CSCG) noise with  mean zero and variance $ \delta^2\bm{I}$. Here, $h_i\in \mathbb{C}$  denotes the Rayleigh channel coefficient \cite{Tse2005}. 
All the channel coefficients are assumed to remain constant during each  transmission block,  but may vary over  blocks. It is assumed that the channel coefficients  and noise variance are perfectly known at both the transmitter and receiver. To overcome fading, \emph{channel inversion} power control  is applied, i.e.,
$
 	p_i=\frac{1}{||h_i||^2}.
$
 As a result, the block fading channel  in \eqref{signal} is  transformed into an AWGN channel, where the received SNRs, $\gamma=\frac{1}{\delta^2}$, are same across  blocks. 

	At the receiver side, the  symbols in the total of $T$ received packets are  decoded into the bit stream $\hat{\bm{b}}$ by the channel decoder.   Then, the bit stream $\hat{\bm{b}}$ is fed into the deep source decoder $\mathcal{F}^{-1}$ to recover the data, i.e., $\hat{\bm{x}}=\mathcal{F}^{-1}(\hat{\bm{b}};\theta_2,\psi)$. The average E2E distortion of the considered system with D$^2$-JSCC can be then defined as 
	\begin{align}\label{e2e_d}
		\mathcal{D}_t=\mathbb{E}_{\bm{x},\bm{N}}\left\{\frac{1}{M}||\bm{x}-\hat{\bm{x}}||^2 \right\},
	\end{align}
	with $\bm{N}=\{\bm{n}_i\}$.

		\subsection{Problem Formulation}
	
	Given the above system model, our goal is to minimize the  
	E2E distortion in \eqref{e2e_d} subject to a constraint on the average channel uses. The optimization problem can be formulated as:
	\begin{align}\label{rate-dis}
		&\ \min_{\{\bm{\Phi},\psi\},R_c} \quad \mathcal{D}_t\\
		&\ \ \ \ \text{s.t.} \quad \quad \mathbb{E}_{\bm{x}}\left\{\left\lceil\frac{B}{L R_c}\right\rceil L\right\}\leq d,  \  R_c\geq 0,
		\nonumber
	\end{align}
	where $d>0$ denotes the maximal number of channel uses. 	When the optimal channel rate, $R_c^*$, for Problem \eqref{rate-dis} is obtained, a $(\lceil L R_c^*\rceil, L)$  block code  can be constructed for  channel coding.    
Using the inequality $\left\lceil \frac{B}{L R_c}\right\rceil L\leq \frac{B}{R_c}+L$, the constraint $\mathbb{E}_{\bm{x}}\left\{\left\lceil \frac{B}{L R_c}\right\rceil L\right\}\leq d$ can be relaxed as $\mathbb{E}_{\bm{x}}\left\{\frac{B}{R_c}\right\}\leq d-L$. Using the relaxation and $R_s=\mathbb{E}_{\bm{x}}\{B\}$,  problem \eqref{rate-dis} becomes
\begin{align} \label{rate-dis2}
		&\ \min_{\{\bm{\Phi},\psi\},R_c} \quad \mathcal{D}_t\\
		&\ \ \ \ \text{s.t.} \quad \quad \frac{R_s}{R_c}\leq \tilde{d},  R_c\geq 0,\nonumber
\end{align}
 where $\tilde{d}=d-L$.

However, there remain several challenges in solving Problem \eqref{rate-dis2}.  First,  
the E2E distortion $\mathcal{D}_t$ has no closed-form expression due to the intractability of NNs, making it  hard to directly optimize the channel rate, $R_c$. Second,  
the quantization operation and the digital source/channel coding  are all discrete functions, which makes it difficult to optimize the parameters $\{\bm{\Phi},\psi\}$ of DNNs by  applying the gradient descent method \cite{dai2022nonlinear,kurka2019}. 


	\section{ Characterization of  E2E Distortion}
	In this section, we characterize the E2E distortion given in \eqref{e2e_d} to help solving Problem \eqref{rate-dis2}. First, we  analyze the Bayesian model of the D$^2$-JSCC system and present some approximations and assumptions on the NNs. Then, based on the approximations, the E2E distortion is characterzied as a function of the NNs parameters and  channel rate.  Lastly, the optimization problem for 	D$^2$-JSCC is reformulated. 
	
	\subsection{Approximations and Assumptions for D$^2$-JSCC}
		We describe in the sequel several approximations to facilitate the development of a approach for designing D$^2$-JSCC, which otherwise is an intractable problem.  First, we  adopt the Bayesian approximation  of  the statistics of the feature  vector $\bm{y}$, its quantized version $\tilde{\bm{y}}$, and the distorted feature vector $\hat{\bm{y}}$. 
		 Recall that in the  deep source encoder, feature $\bm{y}$ conditioned on the latent variables $\tilde{\bm{z}}$ is modeled as an i.n.i.d. Gaussian random variable (see \eqref{norm:1}).  It is common to model the distribution of  $\bm{y}$ as  mixture Gaussian. However, based on our observations in experiments,  most of feature's elements exhibit significant sparsity and their mean values vary slightly w.r.t the latent variables $\tilde{\bm{z}}$. For these reasons, we adopt the following approximations.

				 \begin{figure*}[t]
					\normalsize
					\setlength{\abovecaptionskip}{+0.3cm}
					\setlength{\belowcaptionskip}{-0.1cm}
					\centering
\subfigure[CNN-based model \cite{Balle2018}]{
	\begin{minipage}[t]{\linewidth}
		\centering
		\includegraphics[width=5.in]{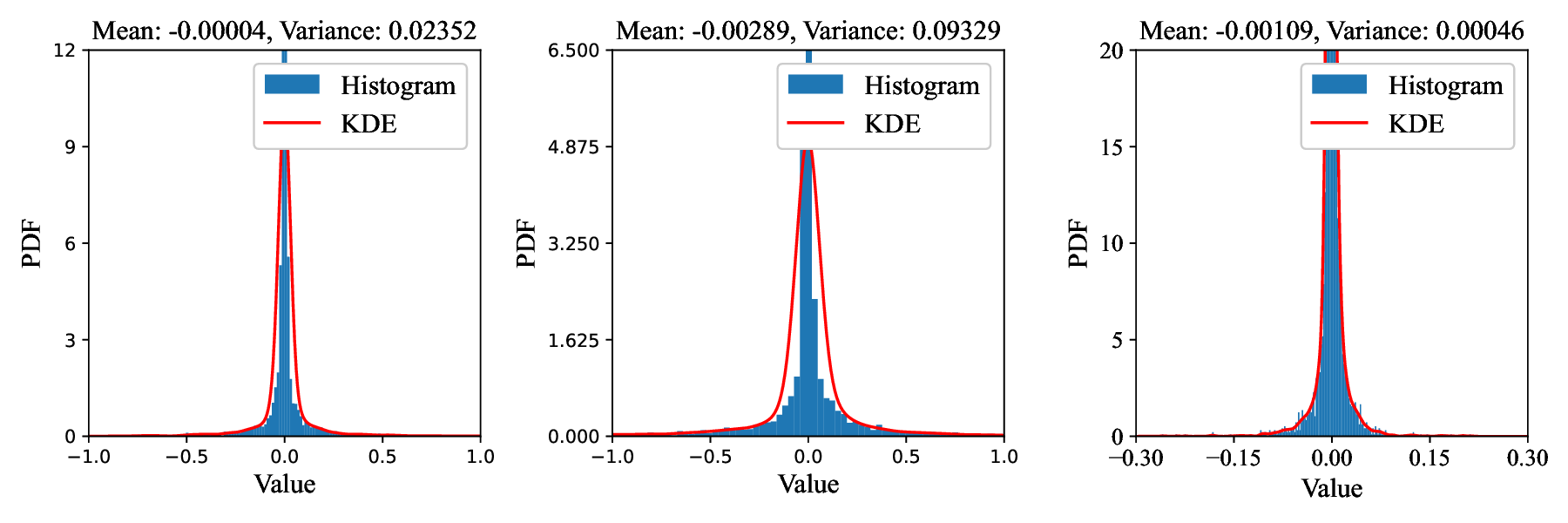}
		
	\end{minipage}%
}%

\subfigure[Transformer-based model \cite{liu2023learned} ]{
	\begin{minipage}[t]{\linewidth}
		\centering
		\includegraphics[width=5.in]{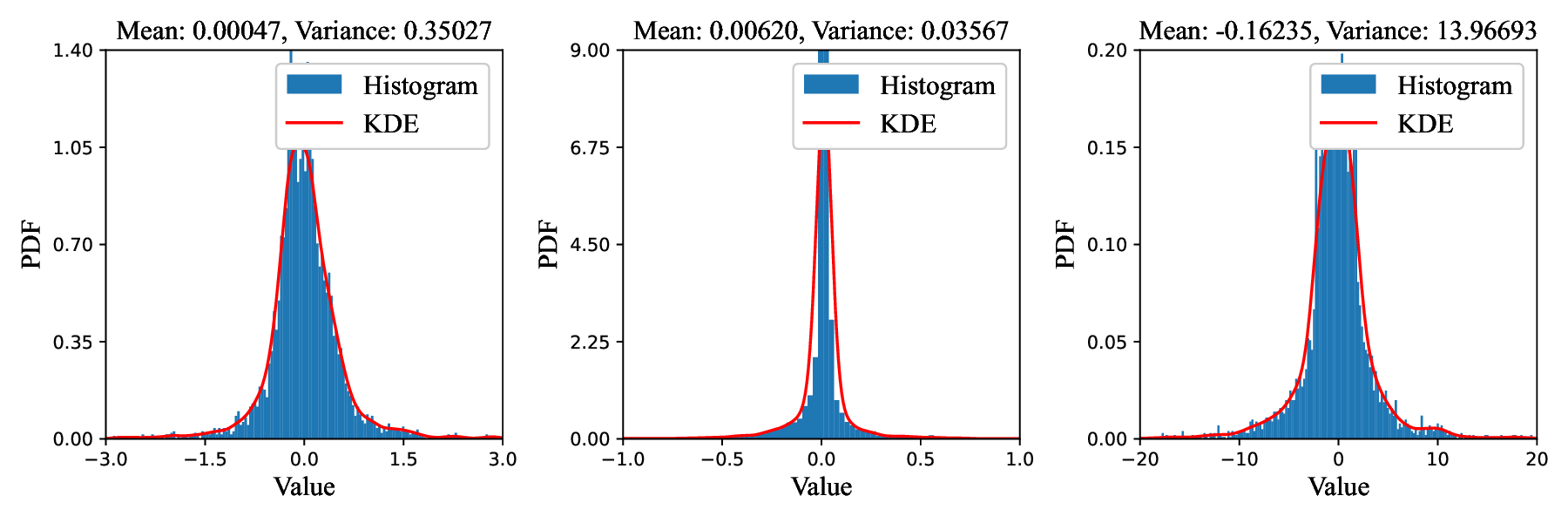}
		
	\end{minipage}%
}%
\captionsetup{justification=justified}
\caption{PDF of feature element $y_i$ with different NN structures. The DNNs are pretrained and the tested images are randomly cropped into $256\times 256$ pixels.  }
\label{assumption1}
				\end{figure*}

				\noindent \textbf{Approximation 1} (Sparse Gaussian approximation).
					\emph{The  feature  $\bm{y}$  is sparse and its elements are described as i.n.i.d.  Gaussian random variables: }
						\begin{align}
							\bm{y} \sim \mathcal{N}(\bar{\bm{u}}_{\bm{\Phi}},\bar{\Sigma}_{\bm{\Phi}} ),
						\end{align}
						\emph{where $\bar{\bm{u}}_{\bm{\Phi}}=[\bar{u}_{\bm{\Phi},1},\bar{u}_{\bm{\Phi},2},\cdots,\bar{u}_{\bm{\Phi},K}]^T$ and $\bar{\Sigma}_{\bm{\Phi}}=\text{\emph{Diag}}(\bar{\sigma}_{\bm{\Phi},1}^2, \bar{\sigma}_{\bm{\Phi},2}^2,\cdots, \bar{\sigma}_{\bm{\Phi},K}^2)$.}
				
\noindent \textbf{Validation:} To validate this approximation, we depict the PDFs of some randomly selected feature elements in Fig. \ref{assumption1}. Without loss of generality, we examined two well-known models: convolutional neural network (CNN) based model \cite{Balle2018} and  transformer-based model \cite{liu2023learned}. The kernel density estimation (KDE) method is utilized to estimate the PDFs of features over a set of samples from a real-world dataset, namely the Open Image Dataset \cite{kuznetsova2020open}. It is observed that for both models, most of the features approximately follow Gaussian distribution with small means and variances, substantiating the assumed feature sparsity.

						\noindent \textbf{Approximation 2} (Uniform  quantization noise).
		\emph{The errors caused by the uniform scalar quantization can be approximated as adding i.i.d. uniform noise into the features vector $\bm{y}$ and the latent-variable vector $\bm{z}$: }
			\begin{align}
				\tilde{\bm{y}}=\bm{y}+\bm{o}_1\  \text{\emph{and}} \  \tilde{\bm{z}}=\bm{z}+\bm{o}_2
			\end{align}
			\emph{where $\bm{o}_1,\bm{o}_2 \sim \mathcal{U}(-\frac{1}{2},\frac{1}{2})$ with $\mathcal{U}(-\frac{1}{2},\frac{1}{2})$  denoting the uniform distribution over the interval $[-0.5,0.5]$. }
			
		\noindent \textbf{Validation}: The uniform approximation is widely adopted in the field of deep image compression, which serves to relax the discrete quantization function and facilitates the application of gradient descent method into training DNNs \cite{Balle2017, Balle2018,9242247}. Furthermore, it's worth noting that the differential entropy of the approximated features, e.g.,  $\tilde{\bm{y}}=\bm{y}+\bm{o}_1$, provides a  positively biased estimate of the discrete entropy of the quantized ones, as discussed in \cite{Balle2017}.  
		
		The distribution of the distorted features in $\hat{\bm{y}}$ are affected by the entropy coding, channel coding, and the SNR. In general, it is  challenging to model the  PDF of  $\hat{\bm{y}}$. The difficulty can be overcome using Approximations 1 and 2 to yield the following result:
		\begin{lemma}\label{Lemma:1}
			 Let  the block error probability be denoted as $\rho$. There exists a constant $\alpha_{\rho,\bm{\Phi}}\geq1$ w.r.t. $\rho$ and $\bm{\Phi}$, such that  the variance of the distorted features $\{\hat{y}_{i}\}$ satisfies
				\begin{align}\label{assum:3}
					\sigma_{\hat{y}_i}^2 \leq \alpha_{\rho,\bm{\Phi}}\left(\bar{\sigma}_{\bm{\Phi},i}^2+\frac{1}{12}\right),\ \forall i,
				\end{align}
				where the equality holds when $\rho=0$ and $\alpha_{\rho,\bm{\Phi}}=1$.
		\end{lemma}
\begin{IEEEproof}
	 The distorted feature vector can be modeled as $\hat{\bm{y}}=\tilde{\bm{y}}+\bm{g}$, where $\bm{g}\in \mathbb{Z}^{K}$ is the perturbations noise caused by channel errors. It is assumed that the error $\bm{g}$ caused by channels is independent of the feature $\tilde{\bm{y}}$. 
	When the block error probability is $0$, it is apparent that the noise $\bm{g}=0$ and the variance of  $\hat{\bm{y}}$ equals to the one of $\tilde{\bm{y}}$. According to Approximations 1 and 2, the variance of the  quantized feature $\tilde{y}_i$ is calculated by  $\bar{\sigma}_{\bm{\Phi},i}^2+\frac{1}{12}$ for all $i \in [K]$.  When the block error probability is larger than $0$,  the channel errors might distort the feature $\tilde{\bm{y}}$, resulting the increasing of variance.  Let the variance of the distorted feature $\hat{y}_i$ be $\sigma_{\hat{y}_i}^2=\alpha_{i}\left(\bar{\sigma}_{\bm{\Phi},i}^2+\frac{1}{12}\right), \alpha_{i} \geq 1, $ and $\alpha_{\rho,\bm{\Phi}}=\max \{\alpha_{1},\alpha_{2},\cdots,\alpha_{K}\}$, and then \eqref{assum:3} is obtained. 
\end{IEEEproof}
		
		\begin{assumption}[Lipschitz continuity]
					The recovery function, $G_{\psi}$, is Lipschitz continuous on $\mathcal{Y}^K$. Specifically,  there exists a positive constant $C_{\psi}$ w.r.t. parameter $\psi$ such that 
			\begin{align}
				||G_{\psi}(\bm{y}_1)-G_{\psi}(\bm{y}_2)||^2\leq C_{\psi}||\bm{y}_1-\bm{y}_2||^2, \  \bm{y}_1,\bm{y}_2\in \mathcal{Y}^K. 
			\end{align}
			The smallest value of  $C_{\psi}$ is called the Lipschitz constant of  $ G_{\psi}$.   
		\end{assumption}

\noindent \textbf{Validation}:  
The Lipschitz constant is usually utilized to measure the sensitivity of the NNs w.r.t the  input perturbations. It can be proved that the commonly used NN layers, such as fully connected and convolutional layers, as well as activation functions like ReLU and Sigmoid are Lipschitz continuous \cite{8054694,LIpstich}. Consequently,  it is reasonable to assume that the NNs, as a composition of these layers and activation functions, are Lipschitz continuous \cite{LIpstich}.

	\subsection{Characterization of E2E distortion}
	We present the main result of the E2E distortion in the following theorem. 
	
	\begin{theorem} \label{Theorem:1}
	Under Assumption 1 and  given the data dimension $M$, there exists a constant $\tilde{\alpha}_{\rho,\bm{\Phi}} > 1$ w.r.t. $\rho,\bm{\Phi}$, such that the E2E distortion given in \eqref{e2e_d} is upper bounded by $\mathcal{D}_t\leq \tilde{\mathcal{D}}_t$ with  $\tilde{\mathcal{D}}_t$ being defined as 
		\begin{align}\label{Theorem1:f}
			 \tilde{\mathcal{D}}_t  & \triangleq              \underbrace{\left( 1-(1-\rho)^{\tilde{T}}\right)\frac{C_{\psi}}{M}(\tilde{\alpha}_{\rho,\bm{\Phi}}-1)\text{\emph{Tr}}\left(\bar{\Sigma}_{\bm{\Phi}}+\frac{1}{12}\bm{I}\right)}_{\text{\emph{Channel distortion}}} \nonumber \\
			 & \quad + \underbrace{\vphantom{\left( 1-(1-\rho)^{\tilde{T}}\right)\frac{C_{\psi}}{M}(1+\alpha_{\rho})\text{Tr}\left(\bar{\Sigma}_{\bm{\Phi}}+\frac{1}{12}\bm{I}\right)}\mathcal{D}_s}_{\text{\emph{Source distortion}}},
		\end{align}	
		where $\tilde{T}=\frac{R_s}{LR_c}$. The equality holds  when $\rho=0$. 
	\end{theorem}
	
	\begin{IEEEproof}
		See Appendix A. 
	\end{IEEEproof}

		From Theorem \ref{Theorem:1}, we have the following observations.
		\begin{enumerate}
			\item It is observed that the upper bound, $\tilde{\mathcal{D}}_t$, consists of two parts: the channel distortion and  the source distortion, $\mathcal{D}_s$, given in \eqref{source-d}. The  former is caused by the block transmission errors.  Specifically, the term $\left( 1-(1-\rho)^{\tilde{T}}\right)$ approximates the average probability of transmitting one data sample in error.  The term, $\left[\frac{C_{\psi}}{M}(\tilde{\alpha}_{\rho,\bm{\Phi}}-1)\text{Tr}\left(\bar{\Sigma}_{\bm{\Phi}}+\frac{1}{12}\bm{I}\right)\right]$,   represents the average penalty when the transmission error occurs.    
			
			\item From \eqref{Theorem1:f},  it is observed  that the channel distortion is a \emph{monotonically increasing} function w.r.t  the block error probability $\rho$, the source rate $R_s$, the Lipschitz constant $C_{\psi}$, and the feature variance. 
This is aligned with the intuition of E2E transmissions, as elaborated in the sequel. First, increasing $R_s$ and $\rho$  result in a higher probability of transmitting a data sample with errors. Next, a larger Lipschitz constant $C_{\psi}$ makes the recovery function $G_{\psi}$ more susceptible to channel errors.  Furthermore, since the distortion is measured by norm-$2$ distance, the penalty caused by transmission error is related to the feature variance.

		\end{enumerate}

To simplify  analysis, we relate  variance,  $\bar{\Sigma}_{\bm{\Phi}}$,  in \eqref{Theorem1:f} with the source rate, $R_s$, and have the following result.
\begin{corollary} \label{Coro:1}
The upper bound on E2E distortion in \eqref{Theorem1:f} can be further upper bounded as $\tilde{\mathcal{D}}_t\leq \hat{\mathcal{D}}_t$  with $\hat{\mathcal{D}}_t$ being defined as 
\begin{small}
	\begin{align}\label{Corr:f1}
	\hat{\mathcal{D}}_t \triangleq 
 \mathcal{D}_s+\frac{K}{M}\left( 1-(1-\rho)^{\tilde{T}}\right)C_{\psi}(\tilde{\alpha}_{\rho,\bm{\Phi}}-1)\left(\frac{2^{2R_s/K}}{2\pi e}+\frac{1}{12}\right).
			\end{align}
			\end{small}
\end{corollary}
\begin{IEEEproof}
	See Appendix B.
\end{IEEEproof}
\begin{remark}\label{Re1}
		Based Theorem \ref{Theorem:1} and Corollary \ref{Coro:1}, we can conclude that minimizing the  source rate, $R_s$, and the block error probability, $\rho$, helps to reduce the channel  distortion caused by transmission errors. However, decreasing  $\rho$ requires a lower channel rate $R_c$, which  increases the bandwidth cost for transmission. On the other hand, the source distortion, $\mathcal{D}_s$,  increases w.r.t. decreasing $R_s$, which can increase the total distortion $\hat{\mathcal{D}}_t$.  Therefore, there exists a trade-off between  $R_c$ and $R_s$. It is important to characterize the trade-off for the purpose of minimizing the E2E distortion subject to a  constraint on the total channel uses.

\end{remark}

\subsection{Problem Approximation}
\begin{figure}[t]
		\centering
		\includegraphics[width=0.9\linewidth]{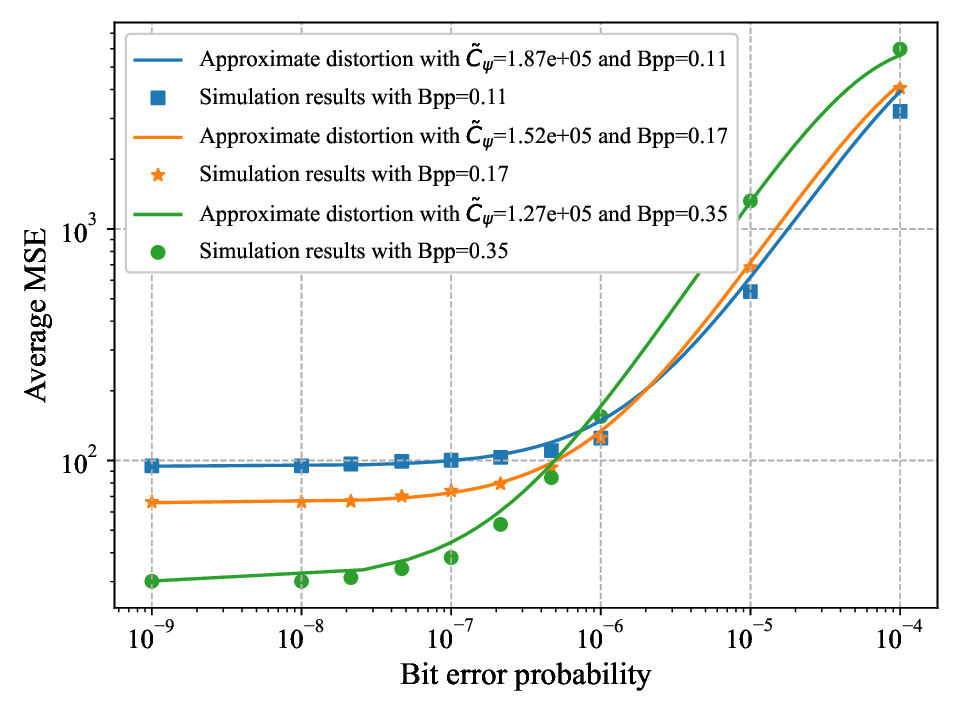}
		\captionsetup{justification=justified}
		\caption{Comparisons of the approximate distortion $\hat{\mathcal{D}}_t$  with the simulation results over different NN models.  The NN structure is CNN-based \cite{Balle2018}.  The tested images are randomly cropped into $256\times 256$ pixels and the Bit per pixel (Bpp) is defined as $\frac{R_s}{256*256}$.  }
		\label{approximationfig}
	\end{figure}

According to Corollary \ref{Coro:1}, minimizing the E2E distortion, $\mathcal{D}_t$, can be relaxed as minimizing its upper bound, $\hat{\mathcal{D}}_t$. However, it is still challenging to accurately characterize $\hat{\mathcal{D}}_t$. One  one hand, the exact Lipschitz constant, $C_{\psi}$,  is hard to estimate \cite{LIpstich}. On the other hand, the parameter, $\tilde{\alpha}_{\rho,\bm{\Phi}}$, which is related to the discrete entropy and channel codings, is hard to express in  closed form.  To address these issues, we let $\left(\tilde{\alpha}_{\rho,\bm{\Phi}}-1\right)$ be a strictly positive constant, denoted as $\hat{\alpha}>0$, and hence $\tilde{C}_{\psi}=\hat{\alpha}C_{\psi}$. Then, we approximate the upper bound, $\hat{\mathcal{D}}_t$, as 
\begin{align} \label{approx:1}
\hat{\mathcal{D}}_t \approx 	\mathcal{D}_s+\mathcal{D}_c,
\end{align}
where  $\mathcal{D}_c$ denotes the channel distortion, defined as
\begin{align}\label{channel:dis}
\mathcal{D}_c=	\frac{K}{M}\left( 1-(1-\rho)^{\tilde{T}}\right)\tilde{C}_{\psi} \left(\frac{2^{2R_s/K}}{2\pi e}+\frac{1}{12}\right).
\end{align}
Here,  the parameter $\tilde{C}_{\psi}$ can be estimated by offline simulations when the NNs parameters $\{\bm{\Phi},\psi\}$  are fixed. 	
	To validate the accuracy of the approximation in  \eqref{approx:1}, the average distortion as a function of the bit error probability, $\rho_b$, is depicted  in Fig. \ref{approximationfig}.  The average block error probability, $\rho$, for a $(N,L)$ block code can be expressed by the bit probability error: $\rho=1-(1-\rho_b)^{N}$.  To simulate the channel environment, we conduct the experiments over Open Image Dataset  \cite{kuznetsova2020open} to calculate the exact MSE $\mathcal{D}_t$ by randomly flipping the encoded bits $\bm{b}$ with the bit error probability $\rho_b$.  From Fig. \ref{approximationfig}, we observe that the approximate distortion, $\hat{\mathcal{D}}_t$, calculated from \eqref{approx:1} match the simulation results well, validating the said approximation. 
	
	Based on the approximate  E2E distortion given in \eqref{approx:1}, the original Problem \eqref{rate-dis2} can be relaxed as:
\begin{align} \label{rate-dis3}
		&\ \min_{\{\bm{\Phi},\psi\},R_c} \quad \hat{\mathcal{D}}_t\\
		&\ \ \ \ \text{s.t.} \quad \quad \frac{R_s}{R_c}\leq \tilde{d}, R_c\geq 0. \nonumber
\end{align} 
According to Remark \ref{Re1}, the key idea of solving  Problem \eqref{rate-dis3} is to find the trade-off between the source rate, $R_s$, and the channel rate, $R_c$. Materializing the idea is challenging, due to the coupling among the parameters $\{\bm{\Phi},\psi, R_c\}$. Moreover, while training the DNNs, the estimation of the parameter $\tilde{C}_{\psi}$ incurs prohibitive computational complexity. 
%
%
%
%
	
	\section{Optimal Rate Control  for D$^2$-JSCC}
	In this section, we propose an efficient algorithm to optimize the source and channel rates for D$^2$-JSCC. To reduce  computational complexity,  the proposed algorithm is designed to have two-step NN-optimization procedure: 1) Source-encoding model selection and 2) model retraining, which are presented in the following subsections.

	\subsection{Step I: Source-encoding Model Selection}
	
	Model selection  in deep learning aims to choose the most appropriate architecture and hyper-parameters of  DNNs for a specific task from  pretrained models \cite{cawley2010over}. It helps  reducing overfitting and is less computationally expensive than training from random initialization. In Step I, we utilize the model-selection method to optimize the DNNs of the deep source encoder and decoder. 	In the followings,  a look-up table with different DNN models is developed and then the joint optimization algorithm for the DNNs and channel rate is presented. 
	
	\textbf{1) Model look-up table:}  The parameters $\{ \bm{\Phi},\psi\}$ of a set of DNNs are trained under  error-free transmissions (i.e., $\tilde{\bm{y}}=\hat{\bm{y}}$). The key idea  is to balance the source rate, $R_s$, and the source distortion, $\mathcal{D}_s$ via minimizing the rate-distortion function  similarly  as in \cite{Balle2018}:
	\begin{align} \label{source-func}
		\min_{\{\bm{\Phi},\psi\}} \lambda \mathcal{D}_s+R_s,
	\end{align} 
	where $\lambda>0$ is a hyper-parameter. Given Approximation 2 of uniform quantization error, the back-propagation method can be easily utilized to train the parameters $\{ \bm{\Phi},\psi\}$ by solving Problem \eqref{source-func}. 
	By varying $\lambda$, a set of DNN models associated with different source rates and distortion  levels can be obtained. For each DNN model, the parameters $\tilde{C}_{\psi}$, $R_s$, and $D_s$
  can be estimated using the validation dataset. Combing the above operations, a look-up table of $P$ models with different source rates, source distortion levels, hyper-parameters $\lambda$, and parameter $\tilde{C}_{\psi}$ is constructed.

	\textbf{2)  Joint model selection and rate control:}  To validate the generality of the proposed algorithm, we consider two types of block channel coding, namely random coding and polar coding. While random coding may not be implementable in practice, it offers a lower bound on the block error probability for finite block length transmissions, which serves as the E2E performance limit of the D$^2$-JSCC \cite{polyanskiy2010channel}. On the other hand, polar codes have been shown to provide excellent error-correcting performance with low decoding complexity for practical block lengths \cite{arikan2009channel}. The relationship between the block error probability and the channel rate, $R_c$, over the equivalent AWGN channel is specified in \eqref{randomc} and \eqref{polarc}.

	Based on the preceding look-up table and the specific block codes, we resort to an efficient algorithm to optimize Problem \eqref{rate-dis3}. 
	When the  NN parameters $\{\bm{\Phi},\psi\}$ are fixed,  Problem \eqref{rate-dis3} reduces to
		\begin{align}\label{subproblem1}
				&\ \min_{R_c} \quad -(1-\rho)^{\tilde{T}} \\
		&\ \ \ \text{s.t.} \quad {R_c}\geq \frac{R_s}{\tilde{d}}. \nonumber	
	\end{align}
	For small $\rho$, $(1-\rho)^{\tilde{T}} \approx (1-\tilde{T}\rho)$. Then, problem \eqref{subproblem1} becomes 
	\begin{align}\label{modelselec:23}
						&\ \min_{R_c} \quad \quad \tilde{T}\rho\\
		&\ \ \ \text{s.t.} \quad {R_c}\geq \frac{R_s}{\tilde{d}}. \nonumber	
	\end{align}
	From Problem \eqref{modelselec:23}, we observe that when the  channel rate is larger than the capacity, i.e., $\frac{R_s}{\tilde{d}}\geq \log_{2}(1+\gamma)$, it is impossible to construct a block code to achieve  reliable communications due to a high error probability ($\rho \approx 1$). Hence, the optimal DNN model for Problem \eqref{rate-dis3} must ensure  $\frac{R_s}{\tilde{d}}\leq \log_2(1+\gamma)$.

	\begin{lemma}\label{Lemma:2}
		Suppose that $\frac{R_s}{\tilde{d}}\leq \log_2(1+\gamma)$. For  random coding and a sufficiently long block length $L$, the optimal solution $R_c^*$ for Problem \eqref{modelselec:23} is given as
		$
			R_c^*=\frac{R_s}{\tilde{d}}.
		$
		For the polar coding  with the block error probability given in \eqref{polarc}, the optimal solution $R_c^*$ of problem \eqref{modelselec:23} is given as 
		\begin{align}\label{lemma2:fuc1}
			R_c^*=\left\{\begin{array}{ll}
				\frac{R_s}{\tilde{d}}, & \text{\emph{if}}\ \ \frac{R_s}{\tilde{d}}\geq\frac{1}{\beta_1},\\
				\frac{1}{\beta_1}, & \text{\emph{Otherwise}}.
			\end{array}\right. 
		\end{align}
			\end{lemma}
	\begin{IEEEproof}
	See Appendix C. 	
	\end{IEEEproof}
	Based on Lemma \ref{Lemma:2},  we utilize the exhaustive search algorithm  to find the best DNN model with $\frac{R_s}{\tilde{d}}\leq \log_2(1+\gamma)$ and  the lowest E2E distortion $\hat{\mathcal{D}}_{t}(\{\bm{\Phi},\psi\},R_c^*)$. The algorithm for the joint model selection and rate optimizations is summarized in Algorithm \ref{Algorithm1}.
Its computational complexity is related to the size of the look-up table, i.e., $O(P)$.

\begin{algorithm}[thb]
		\caption{ Joint Model Selection and Rate Control.}
		\label{Algorithm1}
		\hrule
		\vspace{0.3cm}
		\begin{algorithmic}[1]
			\Require Block length $L$, SNR $\gamma$, block coding type.
			\Ensure $ R_c^*$ and $\{\bm{\Phi},\psi\}^*$.
			
			\State Change the hyper-parameter $\lambda$ and train the NNs by applying back propagation method \cite{lecun2015deep} into  problem \eqref{source-func}. 
			\State Establish the look-up table with different source rate $R_s$, source distortion $\mathcal{D}_s$, hyper-parameter $\lambda$, and constant $\tilde{C}_{\psi}$.
			
			\State For each NN model with parameters $\{\bm{\Phi},\psi\}$ and source rate $R_s$, calculate the distortion $\hat{\mathcal{D}}_t(\{\bm{\Phi},\psi\})$ given in \eqref{approx:1} according to Lemma \ref{Lemma:2} and the block coding type. 
			
			\State Select the best NN model from the look-up table that minimizes $\hat{\mathcal{D}}_t(\{\bm{\Phi},\psi\})$.

            \State  Let $\{\bm{\Phi},\psi\}^*=\{\bm{\Phi},\psi\}^i$, where $i$ is the index of the best NN model. $R_c^*$ is calculated by applying Lemma \ref{Lemma:2}.
		\end{algorithmic}
\end{algorithm}
	
	\subsection{Step II: Source-encoding Model Retraining}
	
	 The preceding model-selection method is merely step $1$ of optimizing the deep source encoder/decoder as the results are sub-optimal for two reasons. The first is the limited number of DNNs in the look-up table, and the second is that the output coders are still independent of the channel SNR.  
	In this subsection, the obtained NN parameters $\{\bm{\Phi},\psi\}$ are retrained to adapt to the channel SNR, thereby approaching   the optimal solution of Problem \eqref{rate-dis3}. 
	
	To this end, we derive a useful result characterizing 
		 the scaling  of the channel distortion $\mathcal{D}_c$  in  \eqref{channel:dis} as the SNR decreases. 
		\begin{theorem}\label{Theorem1:1}
		Given the NN parameters $\{\bm{\Phi},\psi\}$  and the channel rate $R_c$  computed according to Lemma \ref{Lemma:2},  the channel distortion $\mathcal{D}_c$ with random coding\footnote{Since random coding provides a lower bound on the block error probability, the channel distortion with other block codes will increase more quickly than the order given in \eqref{step2:theorem}.} increases in the following order, 
			\begin{align}  \label{step2:theorem}
				\mathcal{D}_c =  O\left(\text{\emph{exp}}\left(-\frac{L(\log_2(1+\gamma)-\frac{R_s}{\tilde{d}})^2}{2\log_2^2(e)}\right)\right),
			\end{align}
			 as $  \log_2(1+\gamma) \rightarrow \left(\frac{R_s}{\tilde{d}}\right)^+$.
		\end{theorem}
		\begin{IEEEproof}
			See Appendix D. 
		\end{IEEEproof}
		
		\begin{remark} \label{Remark: them}
			From Theorem \ref{Theorem1:1}, we can observe that a decrease in the channel capacity $\log_2(1+\gamma)$, will lead to an exponential increase in the channel distortion. According to the properties of exponential functions, the channel distortion approaches zero  in the high SNR regime but  undergos a  surge when the SNR $\gamma$ falls below a certain threshold. This behavior is commonly referred to as the ``cliff effect''.  However, the result in \eqref{step2:theorem} suggests that decreasing the source rate $R_s$ helps to mitigate the cliff effect, as illustrated in the sequel.
		\end{remark}
		
		To substantiate the conclusions  in Remark \ref{Remark: them},  we examine the log inverse of the distortion, i.e., $10 \log (\frac{1}{\hat{\mathcal{D}_t}})$,  w.r.t. the SNR $\gamma$ after Step I algorithm as shown in Fig. \ref{step2fig1}. It is  observed that with the SNR decreasing, the E2E performance will descend in a stepped manner.  This phenomenon is due to the fact that when the channel distortion exponentially increases,  Step I  algorithm  selects another NN model with a lower source rate $R_s$ to suppress the cliff effect.  To further analyze this issue, the behavior of the E2E performance can be described as two stages: ``leveling-off stage'' and ``cliff stage'':

  		\begin{figure}[t]  
	\centering
	\includegraphics[width=3.5in]{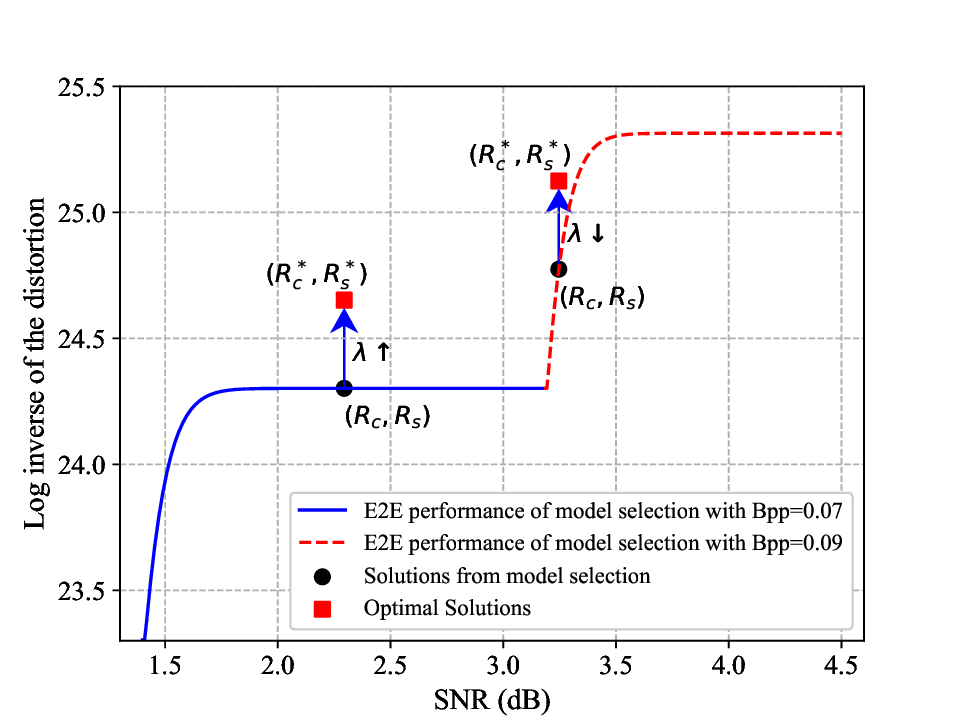}
	\captionsetup{justification=justified}
	\caption{E2E performance of the D$^2$-JSCC system with  the joint model selection and rate control algorithm. The experiments are conduced over Open Image Dataset with random coding,  block length $L=512$, and bandwidth ratio being $0.02$.  }
	\label{step2fig1}
\end{figure}

	\begin{itemize}
		\item ``Leveling-off stage'' denotes the SNR interval  over which the channel distortion  approaches to zero, i.e., $F_1=\{\gamma| 0 \leq \mathcal{D}_c \leq \eta_1  \}$ for small $
		\eta_1 >0$.   In this stage,  the E2E performance is limited by the source encoding, i.e., the source distortion $\mathcal{D}_s$ dominates the E2E distortion $\hat{\mathcal{D}}_t$. To enhance the E2E performance, the NNs need to be retrained to extract more information from  data. In another word, the source rate, $R_s$, needs to be increased to reduce the source distortion, $\mathcal{D}_s$. 
		
		\item ``Cliff stage" denotes the SNR interval over which the channel distortion  dramatically increases, i.e., $F_2=\{\gamma| \mathcal{D}_c \geq \eta_1  \}$ for  $
		\eta_1 >0$.  In this stage, the cliff effect occurs.  According to Theorem \ref{Theorem1:1},  the NNs need to be retrained to reduce the source rate, $R_s$, to mitigate the cliff effect. 
	\end{itemize}

In summary, retraining the DNNs to adaptively control the source rate $R_s$ based on the channel SNR $\gamma$  is crucial for enhancing E2E performance.   An effective method for controlling the DNNs is  to adjust the hyper-parameter $\lambda$ as defined in \eqref{source-func}. It has been noted that training the DNNs with a higher $\lambda$ leads to a consistent decrease in the source distortion, $\mathcal{D}_s$, while simultaneously increasing  $R_s$. Inspired by this idea and the scaling behavior of the channel distortion given in Theorem \ref{Theorem1:1}, we propose an iterative  algorithm  to find the optimal $\lambda^*$ and retrain the DNNs.

 Let $\lambda^{1*}$, $\{\bm{\Phi},\psi\}^{1*}$, $\mathcal{D}_c^{1*}$ and  $\tilde{C}_{\psi}^{1*}$ represent the hyper-parameter, NN parameters, and the estimated parameter from the model selection algorithm, respectively. The NNs are initialized from the parameters $\{\bm{\Phi},\psi\}^{1*}$.  Since the retraining of the NNs focuses  on controlling the source rate, we assume that the estimated parameter $\tilde{C}_{\psi}$ w.r.t. the sensitivity of source decoder stays constant and equals to $\tilde{C}_{\psi}^{1*} $. Then,  the  algorithm for the model retraining iterates the following two steps: 1) model retraining for a fixed hyper-parameter; 2)  hyper-parameter updating, which are elaborated in the following subsections.
 	
 \subsubsection{\textbf{\emph{Model retraining for a fixed hyper-parameter}}}
 By involving  channel distortion into Problem \eqref{source-func}, we propose a new loss function to retrain the DNNs:
		\begin{align}\label{loss:new}
							\hat{\mathcal{L}} =\hat{\lambda}_i \mathcal{D}_s+\beta \hat{\mathcal{D}}_c+R_s,
		\end{align}
		where $\beta>0$ is a constant, and 
\begin{align}\label{new:channel}
\hat{\mathcal{D}}_c=	\frac{K}{M}\left( 1-(1-\rho)^{\tilde{T}}\right)\tilde{C}_{\psi}^{1*} \left(\frac{2^{2R_s/K}}{2\pi e}+\frac{1}{12}\right),
\end{align}
		and the channel rate $R_c$ is calculated from Lemma  \ref{Lemma:2}. 
		 Here, $\hat{\lambda}_i>0, i\geq 1,$ represent the updated hyper-parameter at the $i$-th iteration, which is introduced later.  It is worth noticing that the newly derived loss function $\hat{\mathcal{L}}$ in \eqref{loss:new} differs from that in \eqref{source-func}  by incorporating the channel distortion. The term, $\hat{\mathcal{D}}_c$, therein can be viewed as a regularization term that constrains the increase of the source rate. When the optimal $R_s$ is attained,  $\hat{\mathcal{D}}_c$ approaches zero. In cases where $R_s$ is excessively large, this term helps to decrease it and alleviate the cliff effect.
		 
		The newly derived loss function in \eqref{loss:new}  is differentiable w.r.t. the NN parameters $\{\bm{\Phi},\psi\}$. This allows the back-propagation algorithm to be applied to retraining the DNNs. When the DNNs are well retrained, the source distortion, $\mathcal{D}_{s,i}$, and the source rate,  $R_{s,i}$, can be estimated over the validation dataset. Then, we compute $R_{c,i}$ according to Lemma \ref{Lemma:2}.  Finally, the E2E distortion, $\hat{\mathcal{D}}_{t,i}$, and the channel distortion, $\hat{\mathcal{D}}_{c,i}$,  can be obtained by substituting the solution $(\mathcal{D}_{s,i},R_{s,i},R_{c,i})$ into \eqref{new:channel}.  
		
\subsubsection{\textbf{\emph{Hyper-parameter updating}}}
			The updating of the hyper-parameter, $\hat{\lambda}_{i}$, is based on the scaling behavior of the E2E distortion as shown in Fig. \ref{step2fig1}. At $i$-th iteration, we first determine the stage of the channel distortion $\hat{\mathcal{D}}_{c,i-1}$ after the $(i-1)$-th iteration.
			
			\begin{itemize}
				\item  \textbf{Leveling-off  stage} (i.e., $ 0 \leq \hat{\mathcal{D}}_{c,i-1} \leq \eta_1 $). In this case, the source rate needs to be increased to reduce the source distortion. Initially, $\mathcal{D}_{c,-1}=\mathcal{D}_{c}^{1*}$.   From the look-up table, we can easily find a NN model with a neighbor \footnote{In scenarios where $\lambda^{1*}$ is the largest or smallest one in the loop-up table, we can   set $\tilde{\lambda}$ be a suitable constant larger or smaller than $\lambda^{1*}$. } hyper-parameter $\tilde{\lambda}>\lambda^{1*}$. It is apparent that the optimal hyper-parameter $\lambda^{*}$ for problem \eqref{source-func} falls into the interval $\lambda^{1*} \leq \lambda^{*} \leq \tilde{\lambda}$.  Due to the exponential growth of the channel distortion, the optimal solutions for Problem \eqref{rate-dis3} expect the channel distortion to be small enough compared with the source distortion. Hence, the optimal parameter $\hat{\lambda}$ for problem \eqref{new:channel} is actually similar to the one $\lambda^*$ for problem \eqref{source-func} and also falls into the interval $[\lambda^{1*},\tilde{\lambda}]$.
			Following this idea, we initially  set the hyper-parameter $\hat{\lambda}_0$ as 
			\begin{align}\label{lamba1}
				\hat{\lambda}_0=\frac{\lambda^{1*}+\tilde{\lambda}}{2}.
			\end{align}
The updating of the hyper-parameter $\hat{\lambda}_{i}$  follows the bisection search rule:
\begin{align}
\label{l1}	\hat{\lambda}_{i}=\frac{\hat{\lambda}_{i-1}+\bar{\lambda}_{\max}}{2}, \ \bar{\lambda}_{\min}=\hat{\lambda}_{i-1},
\end{align}
where the parameters $\bar{\lambda}_{\min}$ and $\bar{\lambda}_{\max}$ are initially set as $\text{min}\{\lambda^{1*}, \tilde{\lambda}\}$ and $\text{max}\{\lambda^{1*},\tilde{\lambda}\}$, respectively.

 		\addtolength{\topmargin}{0.01in}	
 		\item  \textbf{Cliff stage} (i.e., $ \hat{\mathcal{D}}_{c,i-1} \geq \eta_1 $). In this stage,  the source rate is reduced, while improving the E2E performance. Similarly, we first find a  neighbor NN model from the look-up table  with the hyper-parameter being smaller than the one obtained from Step I, i.e., $\tilde{\lambda}<\lambda^{1*}$.  The initial hyper-parameter $\hat{\lambda}_0$ is calculated by \eqref{lamba1}.		The updating of the hyper-parameter $\hat{\lambda}_{i}$  follows:
 \begin{align}
\label{l3}	\hat{\lambda}_{i}=\frac{\hat{\lambda}_{i-1}+\bar{\lambda}_{\min}}{2},\ \bar{\lambda}_{\max}=\hat{\lambda}_{i-1},
\end{align}

\end{itemize}
		The updated hyper-parameter $\hat{\lambda}_{i}$ will be used in step 1) for retraining the DNNs.

Finally, alternating the above two steps until the algorithm converges. To summarize, the algorithm for the model retraining is presented in Algorithm 2.

\begin{algorithm}[thb]
		\caption{ Model Retraining}
		\label{Algorithm2}
		\hrule
		\vspace{0.3cm}
		\begin{algorithmic}[1]
			\Require Block length $L$, SNR $\gamma$, block coding type, the parameters $\lambda^{1*}$, $\{\bm{\Phi},\psi\}^{1*}$, $\mathcal{D}_c^{1*}$ and  $\tilde{C}_{\psi}^{1*}$  from Step I, threshold $\eta_1>0$, index $i=0$, tolerance $\xi>0$.
			\Ensure $ R_c^*$ and $\{\bm{\Phi},\psi\}^*$.
					
			\State Determine the stage based on $\mathcal{D}_{c,-1}$ and initialize the hyper-parameter $\hat{\lambda}_0$ according to \eqref{lamba1}.

			\State Initialize the parameters $\bar{\lambda}_{\min}=\text{min}\{\lambda^{1*}, \tilde{\lambda}\}$ and $\bar{\lambda}_{\max}=\text{max}\{\lambda^{1*},\tilde{\lambda}\}$. 
			
			\Repeat 
			\State With the hyper-parameter $\hat{\lambda}_{i}$ and the NN parameters $\{\bm{\Phi},\psi\}^{1*}$, retrain the NNs by using the back propagation method \cite{lecun2015deep} to minimize the loss function $\hat{\mathcal{L}}$ in \eqref{loss:new}. 
			
			\State Calculate the channel rate $R_{c,i}$ according to Lemma \ref{Lemma:2}.
			
			\State Calculate the channel distortion $\hat{\mathcal{D}}_{c,i}$ according to \eqref{new:channel}.

			\If
			 {$\hat{\mathcal{D}}_{c,i} \leq \eta_1 $}:
			 \State Update $\hat{\lambda}_{i}$ and $\bar{\lambda}_{\min}$ according to \eqref{l1}. 
			 \Else
			 \State Update $\hat{\lambda}_{i}$ and $\bar{\lambda}_{\max}$ according to \eqref{l3}. 
			 \EndIf 
			\Until $||\hat{\lambda}_i-\hat{\lambda}_{i-1}||<\xi$; Otherwise, repeat the algorithm and set $i=i+1$. 
			\State $R_c^*=R_{c,i}$ and $\{\bm{\Phi},\psi\}^*=\{\bm{\Phi},\psi\}^i$
		\end{algorithmic}
\end{algorithm}

	\section{Experimental  Results}

\subsection{Experimental Settings}

\begin{itemize}


 	\item \textbf{Model architecture: }
 	 We adopt the classical hyper-prior model in \cite{Balle2018} as the source-encoding architecture to validate the performance gain of the proposed D$^2$-JSCC framework. The hyper-prior model is composed  of convolutional layers with GDN, IGDN, and ReLU activation functions. 
 	 It is worth mentioning that  D$^2$-JSCC framework is also compatible with other NN architectures, e.g., transformer-based architecture \cite{liu2023learned}. 
 	  
 	  	\item \textbf{Real datasets:}  
We test the optimized D$^2$-JSCC system  over two well-known image datasets:    medium-size dataset Kodak  ($768 × 512$ pixels) \cite{Kodak}, and  large-size dataset CLIC  (up to $2048 × 1890$ pixels) \cite{Clic}.  
The dataset for training the deep source coders  in the model-selection step  consists of $100,000$ images sampled from the training dataset of the Open Images Dataset  \cite{kuznetsova2020open}. The look-up table is learned over $10,000$ images randomly sampled from the validation dataset of the Open Images Dataset  \cite{kuznetsova2020open}.  During the model retraining step, we utilize a small portion of the training dataset, i.e.,  $6000$ images, to retrain the deep source encoder/decoder. For  model training and optimization, images are randomly cropped into $256\times 256 $ pixels.
	
	\item \textbf{Model training settings: }In developing the look-up table, we train each NN model for a total of $200$ epochs using the Adam optimizer \cite{Adam} and a mini-batch size of 16. The initial learning rate is set to $10^{-4}$ and is multiplied by $0.1$ when the computed loss remains unchanged. The established look-up table comprises $16$ models with Bpp values ranging from $0.012$ to $1.36$. During  model retraining,  the training epoch for each iteration is set as $10$.  Due to Approximation 2 for quantization noise, the calculated source rate $R_s$ might be larger than the exact one. To better control the hyper-parameter, we first subtract the source rate by a positive constant and then scale the loss functions in  \eqref{source-func} and \eqref{loss:new}
	as follows: $\lambda \mathcal{D}_s+\frac{\tilde{R}_s}{256*256}$ and $\hat{\lambda}_i \mathcal{D}_s+\beta \hat{\mathcal{D}}_c(\tilde{R}_s)+\frac{\tilde{R}_s}{256*256}$, where  $\tilde{R}_s=R_s-0.1*256*256$.  The hyper-parameter $\beta$ is set as $10^{-4}$.  The threshold  $\eta_1$ is set as $1$.
All the experiments were conducted using the PyTorch backend \cite{paszke2019pytorch} on a hardware platform equipped with an Intel(R) Xeon(R) Silver 4210R CPU, NVIDIA A100 GPU, and 40GB of RAM.
 \item \textbf{Channel codes:}
 For channel coding in D$^2$-JSCC, we consider  both the ideal random coding and  the practical polar coding. For the former, we assume that  bit errors uniformly occur over  source bits. Thereby, transmission errors can be simulated by randomly flipping the source bits with a bit error rate of $\rho_b$. The rate is calculated from \eqref{randomc}, and $\rho=1-(1-\rho_b)^{N}$. For polar coding, the achievable channel rate increases as the SNR grows.  This makes it necessary to adapt the modulation type to the channel SNR. To this end, the modulation type is set as BPSK when the SNR is lower than $3$ dB or otherwise, quadrature phase shift keying (QPSK).
Let $b$ denote the modulated bits per symbol (e.g., $b=2$ for QPSK). When the optimal channel rate, $R_c^*$, is achieved, a $(\lceil 4096 \frac{R_c^*}{b}\rceil, 4096)$  polar code  can be constructed for transmission via the aff3ct toolbox \cite{Cassagne2019a}. 

\item \textbf{Benchmark schemes: }
 The benchmark schemes include both the  deep JSCC schemes  \cite{dai2022nonlinear,kurka2019} and the classic separated source-channel coding schemes. For  deep JSCC, we consider two  architectures: the  classic deep JSCC \cite{kurka2019}, namely, DJSCC, and the  nonlinear transform source-channel coding (NTSCC) \cite{dai2022nonlinear}. Both the DJSCC and NTSCC schemes directly employ the DNNs to map image data into analog symbols for transmission, while the latter involves an adaptive density model and demonstrates superior performance. For the separated source-channel coding schemes, we utilize the BPG scheme combined with a $c$-rate $(\lceil4096c\rceil,4096)$ polar code and QPSK modulation. 
 We compare the schemes and D$^2$-JSCC over a block Rayleigh fading channel  with channel-inversion transmission.

\end{itemize}

	\begin{figure}[t]
	\centering
	\subfigure[$\text{SNR}=1$ dB]{
		\begin{minipage}[t]{\linewidth}
			\centering
			\includegraphics[width=3.2in]{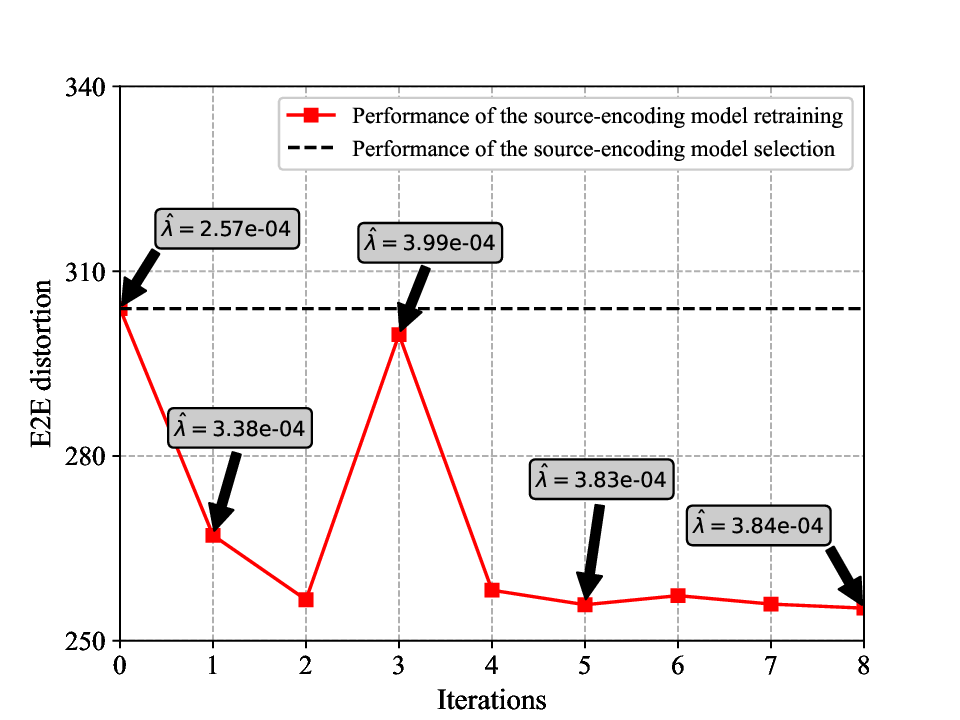}
			
		\end{minipage}%
	}%
	
	\subfigure[$\text{SNR}=7.89$ dB]{
		\begin{minipage}[t]{\linewidth}
			\centering
			\includegraphics[width=3.2in]{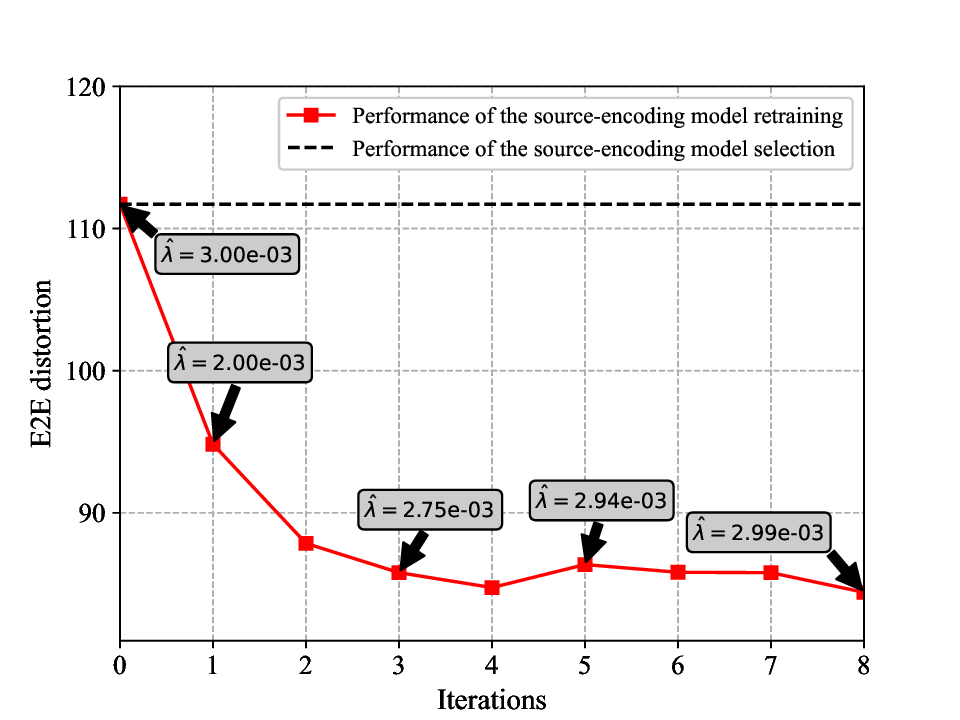}
			
		\end{minipage}%
	}%
	\captionsetup{justification=justified}
	\caption{Convergence performance of the  model retraining algorithm. The average bandwidth ratio is  $0.022$.}
	\label{fig:convege}
\end{figure}

\subsection{Convergence Performance of D$^2$-JSCC}

 Fig. \ref{fig:convege} depicts the E2E distortion as  the number of iteration increases. To better illustrate the convergence performance of the proposed algorithm, we consider two performance stages after the model selection: the leveling-off  stage and the cliff stage. When the SNR equals to $1$ dB, the calculated channel distortion $\mathcal{D}_c$ approaches zero, indicating the leveling-off  stage.   In this stage, the hyper-parameter $\hat{\lambda}$ needs to be increased to enable the NN model to extract more feature information. As shown in Fig. \ref{fig:convege}(a), it is observed that the hyper-parameter gradually increases until it converges to $3.84\times10^{-4}$. It is noted that when the iteration equals $3$, the E2E distortion dramatically increases. This phenomenon can be explained by the fact that when the hyper-parameter $\hat{\lambda}=3.99\times10^{-4}$,  the source rate is too large to be supported by the channel codes with the SNR being 1 dB, leading to the  cliff effect. To decrease the E2E distortion, the proposed algorithm chooses a smaller hyper-parameter $\hat{\lambda}$  to reduce the source rate. A similar phenomenon can be observed in Fig. \ref{fig:convege}(b) with the initial $\mathcal{D}_c\gg 0$. Initially, the hyper-parameter is reduced from $3\times 10^{-3}$ to $2\times 10^{-3}$ to mitigate the cliff effect.  Then,  the hyper-parameter  gradually increases to reach the optimal value. In addition, Fig. \ref{fig:convege} reveals  that  the proposed retraining algorithm quickly converges and achieves a significant performance gain compared with the model selection algorithm.

		\begin{figure}[t]
	\centering
	\subfigure[Kodak dataset]{
		\begin{minipage}[t]{\linewidth}
			\centering
			\includegraphics[width=3.2in]{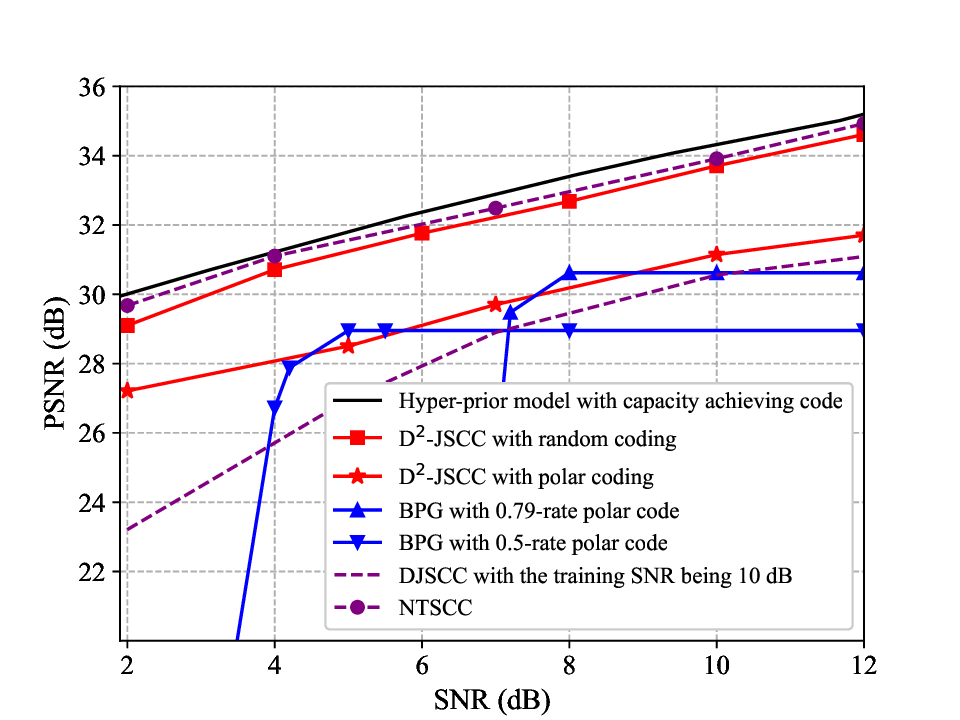}
			
		\end{minipage}%
	}%
	
	\subfigure[CLIC dataset ]{
		\begin{minipage}[t]{\linewidth}
			\centering
			\includegraphics[width=3.2in]{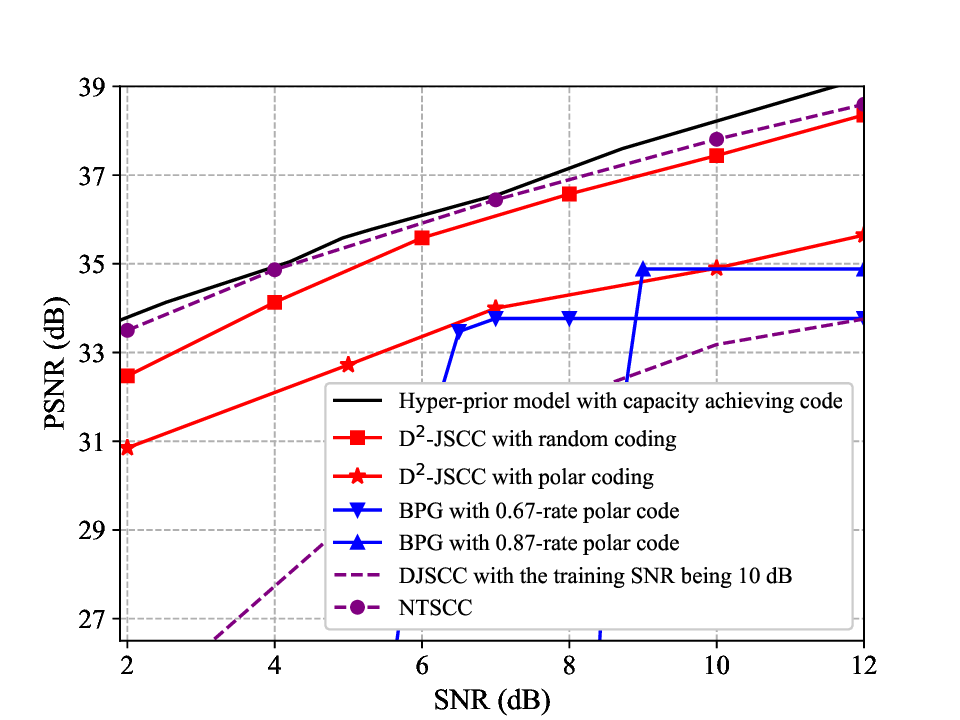}
			
		\end{minipage}%
	}%
	\captionsetup{justification=justified}
	\caption{PSNR performance versus the SNR over different datasets. The average bandwidth ratio is  set as  $0.0625$ and the block length for the D$^2$-JSCC scheme with random coding is $1024$. }
	\label{fig:PSNR1}
\end{figure}

\subsection{E2E Performance of D$^2$-JSCC}
 First, we quantify the E2E performance of the proposed D$^2$-JSCC scheme using  the widely used pixel-wise metric, i.e., the peak signal-to-noise ratio (PSNR) \cite{Balle2017,Balle2018}. Fig. \ref{fig:PSNR1} depicts the PSNR performance as a function of the SNR over different datasets.  The ``hyper-prior model with capacity-achieving code" can be seen as a performance bound on using the deep source coding, where the encoded bits are transmitted error-free at the capacity rate. 
For both the Kodak and CLIC datasets, we can observe that the proposed D$^2$-JSCC scheme mitigates the cliff effect and has a significant performance gain compared to the separate source-channel coding across low to high SNR regions. More specifically, as the SNR decreases, the performance of the BPG scheme with a fixed rate exponentially decreases, while the proposed scheme still maintains graceful performance degradation.  For example,  at an SNR of $2$ dB, the proposed D$^2$-JSCC scheme with polar coding still achieves $27.2$ dB and $30.8$ dB over Kodak and CLIC datasets, respectively. 
Additionally, as the SNR increases, the performance of the BPG scheme remains unchanged, while the proposed scheme is able to support image transmission with a higher PSNR. For instance, a $2.7$ dB performance gain can be observed at an SNR of $12$ dB for the Kodak dataset, compared to the BPG scheme with a $0.5$-rate polar code. The reason for these phenomenons is that  the proposed scheme can efficiently balance the source and channel rates to adapt to the variations of the SNR.  

\begin{figure}[t]
	\centering
	\subfigure[Kodak dataset]{
		\begin{minipage}[t]{\linewidth}
			\centering
			\includegraphics[width=3.2in]{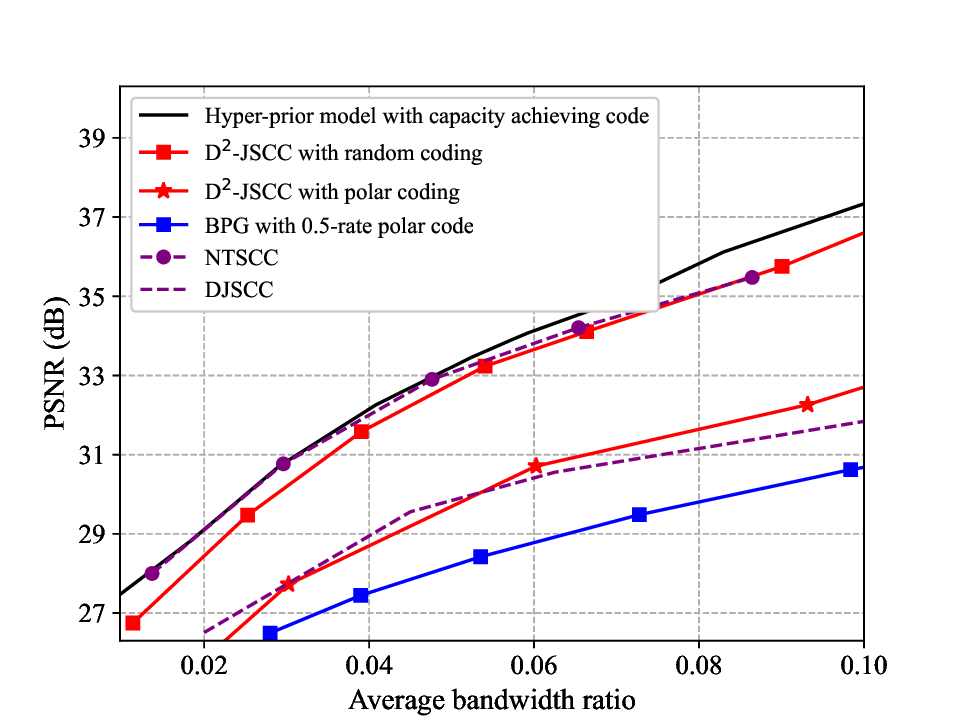}
			
		\end{minipage}%
	}%
	
	\subfigure[CLIC dataset ]{
		\begin{minipage}[t]{\linewidth}
			\centering
			\includegraphics[width=3.2in]{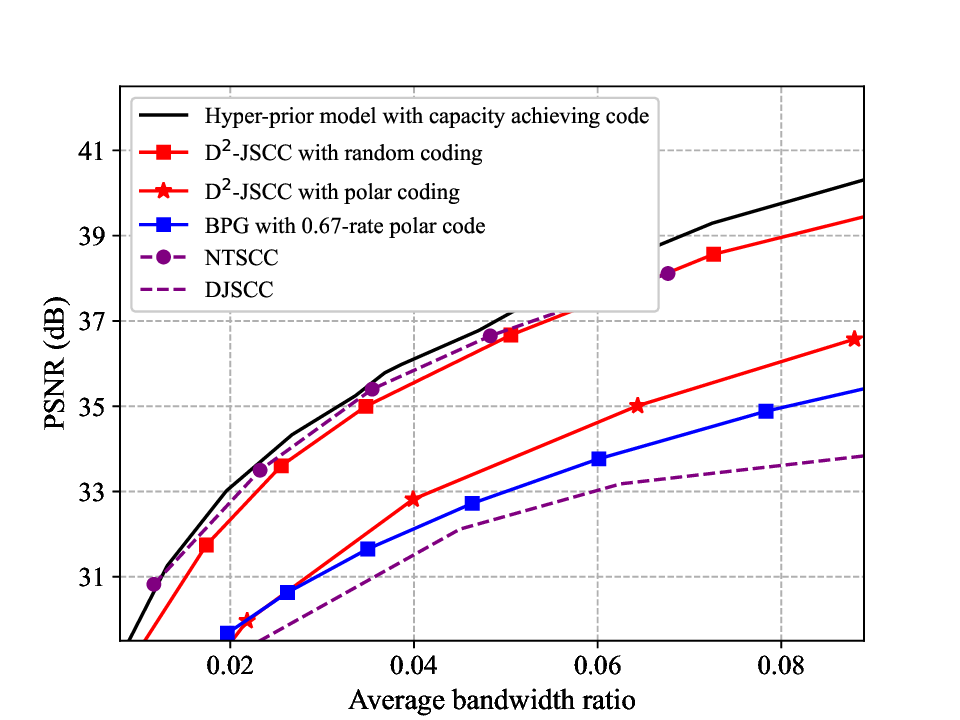}
			
		\end{minipage}%
	}%
	\captionsetup{justification=justified}
	\caption{PSNR performance versus the bandwidth ratio over different datasets. The SNR is set as $10$ dB.}
	\label{fig:ratio1}
\end{figure}

	  		\begin{figure}[t]  
	\centering
	\includegraphics[width=3.2in]{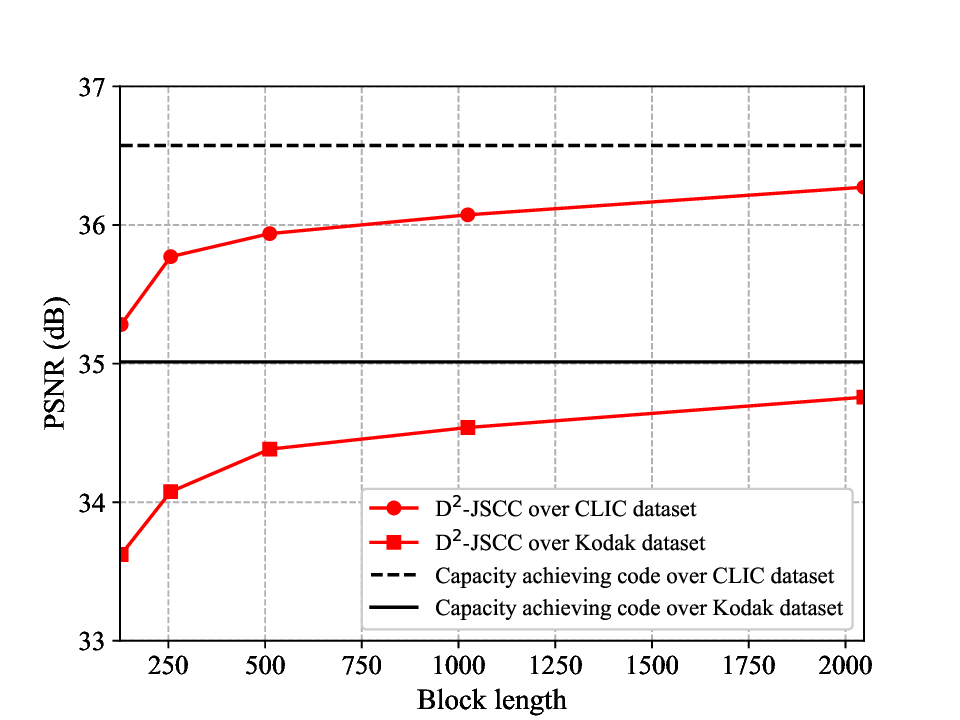}
	\captionsetup{justification=justified}
	\caption{PSNR performance versus the block length over different datasets. The ideal random coding is adopted and the SNR is set as $7$ dB. The bandwidth ratios are  $0.061$ and $0.096$ for CLIC and Kodak datasets, respectively. }
	\label{step44}
\end{figure}

Moreover, we observe from Fig. \ref{fig:PSNR1} that the proposed scheme with random coding has a comparable PSNR performance compared with the NTSCC scheme. For instance, the proposed D$^2$-JSCC scheme with random coding  exhibits only a $0.39$ dB degradation  compared with the NTSCC scheme  for Kodak dataset at the SNR of $4$ dB.  The performance loss comes from the discrete errors caused by quantization, and digital source and channel codes. However, the proposed scheme is more compatible with the current digital communication protocols.  In addition, the optimizations of the proposed scheme do not rely on the instantaneous channel samples but the channel SNR, which makes it more easily implemented in reality. More strikingly, we observe that the proposed scheme has a better performance compared with the DJSCC scheme over some SNR regions. For example, at an SNR of $12$ dB and CLIC dataset, an around $2$ dB gain can be observed. The reason for the performance gain is that the DJSCC fixes the number of transmitted symbols for all images, while the proposed scheme with the adaptive model is able to  change the source-encoded bits based on the content of the images, resulting in a better performance over large-size 
images. 

Next,  in Fig. \ref{fig:ratio1}, we depict the PSNR performance as a function of the bandwidth ratios. It is observed that the proposed scheme exhibits a significant performance gain compared with the BPG scheme across different bandwidth ratios, indicating that the proposed scheme is capable of saving more bandwidths while maintaining the same PSNR.  For instance, when the dataset is Kodak and the PSNR is $31$ dB, the proposed scheme with polar coding is able to save around $0.04*M$ bandwidths compared with the BPG scheme.  When the dataset is CLIC, to achieve $35$ dB PSNR, the proposed scheme is able to save around $0.015*M$ bandwidths.

We also investigate the performance of the proposed scheme w.r.t. the block length. Fig. \ref{step44} depicts the PSNR performance as a function of the block length over different datasets. With the block length increasing, the performance of the proposed scheme will approaches to the performance bound with the capacity achieving code. For example, when the block length is $2048$ and the dataset is Kodak, performance gap between the proposed scheme and the bound is around $0.25$ dB. The reason for this phenomenon  is that with the block length increasing, the achievable rate for the reliable image transmission will increase to the capacity.  This phenomenon is aligned with the  Shannon theorem  that the separate source-channel coding is optimal when the block length tends to infinity \cite{gallager1968information}. 

	\begin{figure}[t]
	\centering
	\subfigure[Kodak dataset]{
		\begin{minipage}[t]{\linewidth}
			\centering
			\includegraphics[width=3.2in]{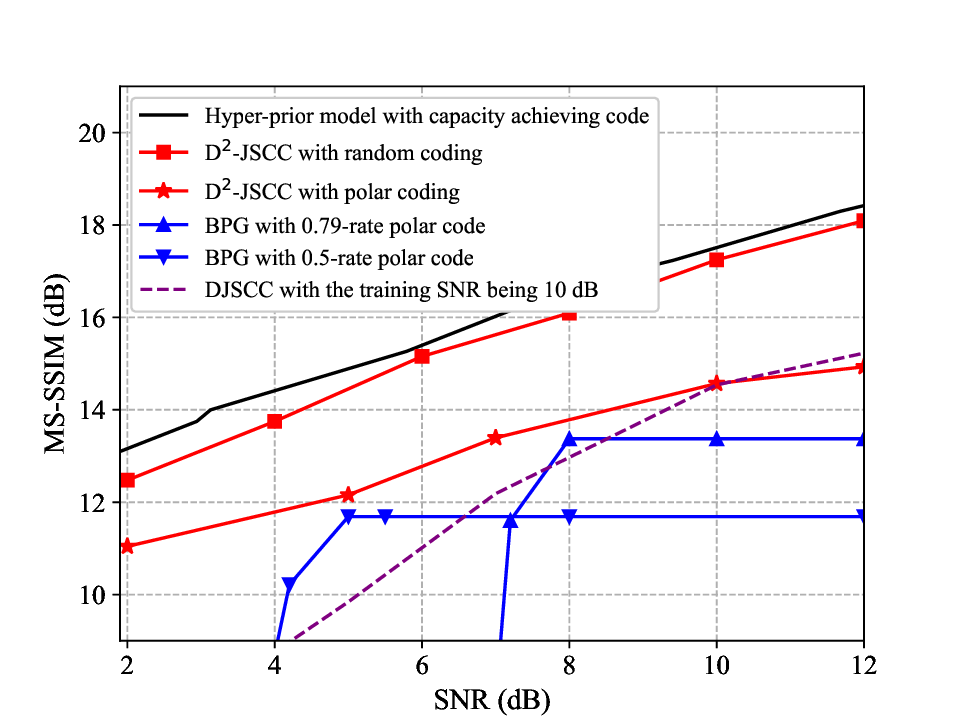}
			
		\end{minipage}%
	}%
	
	\subfigure[CLIC dataset ]{
		\begin{minipage}[t]{\linewidth}
			\centering
			\includegraphics[width=3.2in]{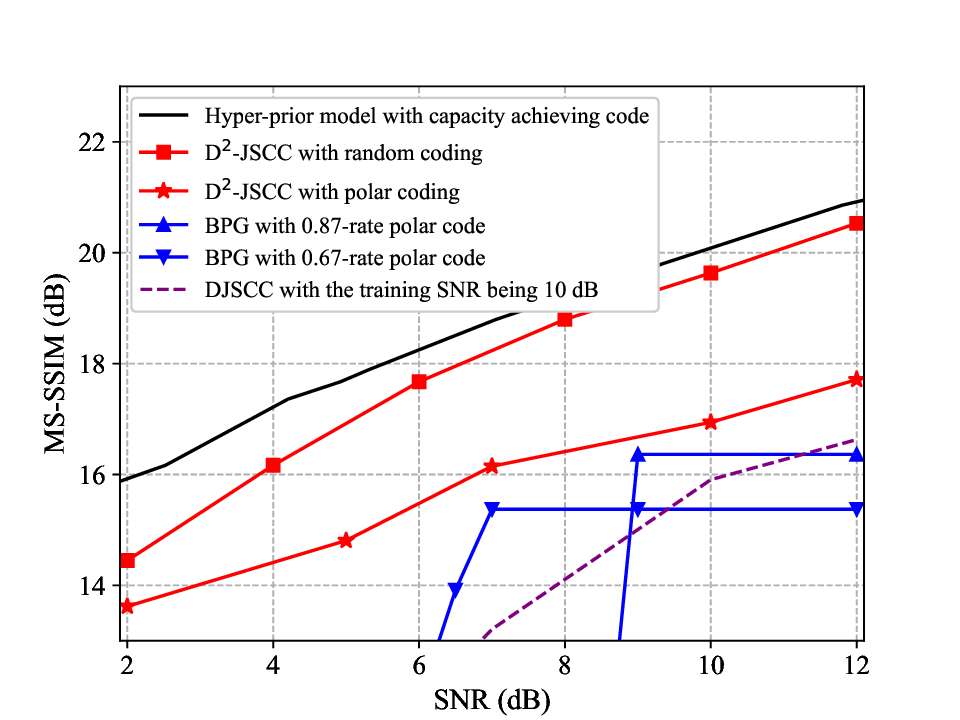}
			
		\end{minipage}%
	}%
	\captionsetup{justification=justified}
	\caption{MS-SSIM performance versus the SNR over different datasets. The average bandwidth ratio is  set as  $0.0625$.}
	\label{fig:PSNR2}
\end{figure}

\subsection{Perceptual Performance of D$^2$-JSCC}

Finally, we evaluate the performance of the proposed scheme using  the perceptual metric, i.e., the multi-scale structural similarity index (MS-SSIM) \cite{wang2003multiscale}. The MS-SSIM metric  is widely used to measure the structural errors of images, rather than the pixel-level distortion in PSNR metric, and is more suitable for quantifying the visual quality of images.  As shown in Fig. \ref{fig:PSNR2}, we depicts the MS-SSIM as a function of the SNR over different datasets. 
It is observed that the proposed scheme still achieves  higher MS-SSIM scores compared with the DJSCC and BPG schemes across different SNR regimes. For instance, when the dataset is Kodak and the SNR is $7$ dB, the proposed scheme with polar coding is $1.8$ dB better than the BPG scheme with a $0.79$-rate polar code.  Here, we do not compare the proposed scheme with the NTSCC, because the latter is able to train the neural networks using the MS-SSIM as the loss function and achieves better results. However, it is possible for the proposed scheme to modify the  distortion function in equation \eqref{approx:1} to improve the MS-SSIM, which will be left for future research.

	%
	
	\section{Concluding Remarks}
This paper proposed a novel D$^2$-JSCC framework for image transmission problem in SemCom, where the digital source and channel coding are jointly optimized to reduce the E2E distortion. Specifically, we designed a deep source coding scheme with an adaptive density model to  encode semantic features according to their different distributions, leading to an increased coding efficiency.  To facilitate the joint design of the source and channel coding, the E2E distortion was characterized as a function of the NN parameters and channel rate.  To minimize the E2E distortion, we proposed a  two-step algorithm with low computational complexity.   Simulation results reveal that the proposed D$^2$-JSCC outperforms both the  classic deep JSCC  and the classical separation-based approaches.  

Some potential research directions on developing the  D$^2$-JSCC framework for SemCom are summarized as follows: 
\begin{itemize}
	\item \textbf{E2E Metric Design}: This paper described the E2E performance using the classical MSE metric, which might not be optimal for certain task-oriented applications. More metric designs(e.g., MS-SSIM and task accuracy) can be explored in the D$^2$-JSCC framework.
	\item \textbf{D$^2$-JSCC with Digital  Communication Techniques}: Future research could explore the integration of the D$^2$-JSCC framework with existing digital communication techniques, such as orthogonal frequency division multiplexing and multiple-input and multiple-output.
	
\end{itemize}

\appendices
\section{Proof of Theorem \ref{Theorem:1}}

According to the considered point-to-point channel given in  \eqref{signal}, the encoded bits  of data $\bm{x}$ is divided into $T$ packets for transmissions. Based on the $(N,L)$ block code and its block error probability $\rho$, we define the probability event $A=\{\text{All the $T$ packets are successfully decoded}\}$ and its complementary event $\tilde{A}$ with $\text{Pr}(A)=(1-\rho)^T$ and $\text{Pr}(\tilde{A})=1-\text{Pr}(A)$. Given the data dimension $M$,  the E2E distortion in \eqref{e2e_d} is expressed as

\begin{align}
	\mathcal{D}_t&=\frac{1}{M}\mathbb{E}_{\bm{x},\bm{N}}\{||\bm{x}-\hat{\bm{x}}||^2 \}, \nonumber \\
	&\overset{(a)}=\frac{1}{M} \mathbb{E}_{\bm{x}} \Big\{ \text{Pr}(A)\mathbb{E}_{\bm{N}}\{||\bm{x}-\hat{\bm{x}}||^2| {A}\} \nonumber\\
	 &\quad \quad \quad  \quad \quad +\text{Pr}(\tilde{A}) \mathbb{E}_{\bm{N}}\{||\bm{x}-\hat{\bm{x}}||^2| \tilde{A}\}       \Big\}, \nonumber \\
	&\overset{(b)}= \frac{1}{M}\mathbb{E}_{\bm{x}} \Big\{ (1-\rho)^{T}\mathbb{E}_{\bm{N}}\{||\bm{x}-\hat{\bm{x}}||^2| {A}\} \nonumber \\
	&\quad \quad \quad  \quad \quad  +\left(1-(1-\rho)^{T} \right)  \mathbb{E}_{\bm{N}}\{||\bm{x}-\hat{\bm{x}}||^2| \tilde{A}\}       \Big\}, \nonumber \\
	&\overset{(c)} \approx(1-\rho)^{\tilde{T}}\frac{1}{M}\mathbb{E}_{\bm{x},\bm{N}}\{||\bm{x}-\hat{\bm{x}}||^2| {A}\} \nonumber \\
	&\quad  +\frac{1}{M}\left(1-(1-\rho)^{\tilde{T}} \right) \mathbb{E}_{\bm{x},\bm{N}}\{||\bm{x}-\hat{\bm{x}}||^2| \tilde{A}\}, \nonumber \\
\label{proof:1}	&\!\overset{(d)}=\!(1-\rho)^{\tilde{T}}\mathcal{D}_s +\left(1-(1-\rho)^{\tilde{T}} \right)\underbrace{\frac{1}{M} \mathbb{E}_{\bm{x},\bm{N}}\{||\bm{x}-\hat{\bm{x}}||^2| \tilde{A}\}}_{\mathcal{K}},
\end{align}
where equality $(a) $ holds due to the  law of total expectation \cite{rohatgi2015introduction }. To prove $(c)$,  we first express  the function  $(1-\rho)^{T}$ as a Taylor series w.r.t. $T$ around the average packet number $\tilde{T}=\frac{R_s}{LR_c}$, i.e., 
$
	(1-\rho)^{T}=(1-\rho)^{\tilde{T}}+\log(1-\rho)(1-\rho)^{\tilde{T}}(T-\tilde{T})+o((\log(1-\rho))(T-\tilde{T})).
$
Assume that the variation of  number of packets is bounded, i.e., $|T-\tilde{T}|<g, g>0$. When the block error probability is small enough, we have $\log(1-\rho)\approx 0$ and $(1-\rho)^{T}$ can be approximated by $(1-\rho)^{\tilde{T}}$. By applying $(1-\rho)^{T}\approx (1-\rho)^{\tilde{T}}$ into (b), (c) is obtained.  Equality $(d)$ holds due to the fact that the term $	\frac{1}{M}\mathbb{E}_{\bm{x},\bm{N}}\{||\bm{x}-\hat{\bm{x}}||^2| {A}\}$ in (c) follows:
\begin{align}
	\frac{1}{M}\mathbb{E}_{\bm{x},\bm{N}}\{||\bm{x}-\hat{\bm{x}}||^2| {A}\}=	\frac{1}{M}\mathbb{E}_{\bm{x}}\{||\bm{x}-\bar{\bm{x}}||^2\}=\mathcal{D}_s, 
\end{align}
where $\bar{\bm{x}}=\mathcal{F}^{-1}(\mathcal{F}(\bm{x}))$ denotes the recovered data under  error-free transmissions.

Then, we bound the term $\mathcal{K}$  given in \eqref{proof:1} as follows:
\begin{align}
	&\frac{1}{M} \mathbb{E}_{\bm{x},\bm{N}}\{||\bm{x}-\hat{\bm{x}}||^2| \tilde{A}\} \nonumber \\
	&\overset{(a)} = \frac{1}{M}\mathbb{E}_{\bm{x},\bm{N}}\{||\bm{x}-\bar{\bm{x}}+\bar{\bm{x}}-\hat{\bm{x}}||^2| \tilde{A}\},\nonumber \\
	&\overset{(b)} =			  \frac{1}{M}\Big( \mathbb{E}_{\bm{x},\bm{N}}\{||\bm{x}-\bar{\bm{x}}||^2| \tilde{A}\} +  \mathbb{E}_{\bm{x},\bm{N}}\{||\hat{\bm{x}}-\bar{\bm{x}}||^2| \tilde{A}\}  \nonumber \\
	& \quad \quad \quad \quad  -2 \mathbb{E}_{\bm{x},\bm{N}}\{(\hat{\bm{x}}-\bar{\bm{x}})^T(\bm{x}-\bar{\bm{x}})| \tilde{A}\}
	\Big),\nonumber \\
	 &\overset{(c)} = \mathcal{D}_s+ \frac{1}{M}\mathbb{E}_{\bm{x},\bm{N}}\{||\bar{\bm{x}}-\hat{\bm{x}}||^2| \tilde{A}\},\nonumber \\
	 &\overset{(d)} = \mathcal{D}_s+ \frac{1}{M}\mathbb{E}_{\bm{x},\bm{N}}\{||G_{\psi}(\tilde{\bm{y}})-G_{\psi}(\hat{\bm{y}})||^2| \tilde{A}\},\nonumber \\
	&\overset{(e)} \leq \mathcal{D}_s+ \frac{1}{M}C_{\psi}\mathbb{E}_{\bm{x},\bm{N}}\{||\tilde{\bm{y}}-\hat{\bm{y}}||^2| \tilde{A}\},\nonumber \\
\label{Theorem: p2}	&\overset{(f)} \leq \mathcal{D}_s+ \frac{1}{M}C_{\psi}(\tilde{\alpha}_{\rho,\bm{\Phi}}-1)\text{Tr}\left(\bar{\Sigma}_{\bm{\Phi}}+\frac{1}{12}\bm{I}\right).
\end{align}
The above equalities and inequalities $(a)$-$(f)$ are proved as follows:
\begin{itemize}
	\item Equality $(b)$ can be obtained by expanding the norm expression in $(a)$.
	\item Equality $(c)$ holds due to the facts that $\frac{1}{M} \mathbb{E}_{\bm{x},\bm{N}}\{||\bm{x}-\bar{\bm{x}}||^2| \tilde{A}\}=\mathcal{D}_s$
	and 
	\begin{align} 
		&\mathbb{E}_{\bm{x},\bm{N}}\{(\hat{\bm{x}}-\bar{\bm{x}})^T(\bm{x}-\bar{\bm{x}})| \tilde{A}\}\nonumber \\
	\label{proof:sub1}	&=\mathbb{E}_{\bm{x},\bm{N}}\{(\hat{\bm{x}}-\bar{\bm{x}})^T| \tilde{A}\}\mathbb{E}_{\bm{x}}\{(\bm{x}-\bar{\bm{x}})\}=0.
	\end{align}
	Equality \eqref{proof:sub1} comes from the facts that the channel errors are independent from the source data and the source encoder is designed to meet the centroid condition \cite{quantization,Balle2018}, i.e., $\mathbb{E}_{\bm{x}}\{(\bm{x}-\bar{\bm{x}})\}=0$.  
	\item Equality $(d)$ is obtained by involving the recovery function $G_{\psi}$ into $(c)$.
	
	\item Inequality $(e)$ is obtained by applying Assumption 3.

	\item To prove inequality $(f)$, we first assume that $\hat{\bm{y}}=\tilde{\bm{y}}+\bm{g}$, where $\bm{g}=[g_1,g_2,\cdots,g_{K}]^T$ is the error caused by transmissions with zero mean, i.e., $\mathbb{E}(g_i)=0$, and is uncorrelated with the feature $\tilde{\bm{y}}$ from source coding. By applying Lemma \ref{Lemma:1}, we have 
	\begin{align}
		\mathbb{E}\{||\bm{g}||^2\}\leq (\alpha_{\rho,\bm{\Phi}}-1)\text{Tr}\left(\bar{\Sigma}_{\bm{\Phi}}+\frac{1}{12}\bm{I}\right).
	\end{align}
	Conditioned on the probability event $\tilde{A}$, we have $\bm{g}\neq0$ and there exist a constant  $\tilde{\alpha}_{\rho,\bm{\Phi}}>\alpha_{\rho,\bm{\Phi}}>1$  such that 
	\begin{align}\label{subproof:2}
				\mathbb{E}\{||\bm{g}||^2|\tilde{A}\}\leq (\tilde{\alpha}_{\rho,\bm{\Phi}}-1)\text{Tr}\left(\bar{\Sigma}_{\bm{\Phi}}+\frac{1}{12}\bm{I}\right).
	\end{align}
By applying \eqref{subproof:2} into $(e)$, we have
	\begin{align}
		\mathbb{E}_{\bm{x},\bm{N}}\{||\tilde{\bm{y}}-\hat{\bm{y}}||^2| \tilde{A}\}\leq (\tilde{\alpha}_{\rho,\bm{\Phi}}-1)\text{Tr}\left(\bar{\Sigma}_{\bm{\Phi}}+\frac{1}{12}\bm{I}\right).
	\end{align}
	Then, inequality $(f)$ is obtained. 
\end{itemize}

Finally, by substituting \eqref{Theorem: p2} into \eqref{proof:1}, Theorem \ref{Theorem:1} is proved. 

\section{Proof of Corollary \ref{Coro:1}}

According to Approximation 2 in Section III, we approximate the quantized feature as $\tilde{\bm{y}}=\bm{y}+\bm{o}$ with $\bm{o} \sim \mathcal{U}(-\frac{1}{2},\frac{1}{2})$. From the property of conditional entropy \cite{gallager1968information}, the entropy of feature $\tilde{\bm{y}}$ follows  $R_s =	H(\tilde{\bm{y}},\tilde{\bm{z}})\geq H(\tilde{\bm{y}}) \geq H(\tilde{\bm{y}}|\bm{o})$ with $H(\tilde{\bm{y}}|\bm{o})$ being calculated as 
\begin{align}
	H(\tilde{\bm{y}}|\bm{o})&=\int_{\bm{o}} \left(\int_{\tilde{\bm{y}}}-p(\tilde{\bm{y}}|\bm{o})  \log p(\tilde{\bm{y}}|\bm{o}) d\tilde{\bm{y}}    \right)d \bm{o}\nonumber\\
	\label{appB:proof1}&=\frac{1}{2}\log_{2}(2\pi e)^{K}+\frac{1}{2}\log_{2}\text{det}(\bar{\Sigma}_{\bm{\Phi}}), 
\end{align}
It is worth noticing that the side information $\tilde{\bm{z}}$ contains significantly fewer bits than the feature $\tilde{\bm{y}}$, which implies that the inequality $R_s \geq H(\tilde{\bm{y}})$ in equation \eqref{appB:proof1} is indeed tight. 
 From \eqref{appB:proof1}, we have 
\begin{align}
	\label{AppB:2} 2^{2R_s-\log_{2}(2\pi e)^K} &\geq \text{det}(\bar{\Sigma}_{\bm{\Phi}})
	 \approx \left(\frac{1}{K}\text{Tr}(\bar{\Sigma}_{\bm{\Phi}})\right)^{K},
\end{align}
where approximation \eqref{AppB:2} comes from Approximation 1 that feature $\bm{y}$ is sparse and most of variances $\{\bar{\sigma}_{\bm{\Phi},i}^2\}$ are small and similar.  Then, we have 
\begin{align}\label{AppB:3}
	\text{Tr}(\bar{\Sigma}_{\bm{\Phi}})\leq K\frac{2^{2R_s/K}}{2\pi e}.
\end{align}
By substituting \eqref{AppB:3} into \eqref{Theorem1:f}, Corollary \ref{Coro:1} is proved. 

\section{Proof of Lemma \ref{Lemma:2}}
For the random coding and ML decoder with the block error probability given in \eqref{randomc}, the objective function in problem \eqref{modelselec:23} becomes
\begin{align}
	U_{r}=\frac{R_s}{LR_c}Q\left(\frac{\sqrt{L}\left(\log_2(1+\gamma)-R_c\right)}{\sqrt{\left(1-\frac{1}{(1+\gamma)^2}\right)\log_2^2(e)}}\right).
\end{align}
Let $a=\frac{\sqrt{L}\log_2(1+\gamma)}{\sqrt{\left(1-\frac{1}{(1+\gamma)^2}\right)\log_2^2(e)}}$ and $b=\frac{\sqrt{L}}{\sqrt{\left(1-\frac{1}{(1+\gamma)^2}\right)\log_2^2(e)}}$
. The first order gradient of function $U_r$ w.r.t. $R_c$ is calculated by
\begin{small}
\begin{align}
	\frac{\partial U_{r}}{\partial R_c}&=\frac{R_sb}{LR_c}\frac{1}{\sqrt{2\pi}}\text{exp}\left(-\frac{(a-bR_c)^2}{2}\right)-\frac{R_s}{LR_c^2}Q(a-bR_c),\\
\label{lemma21}	 &\geq \frac{R_sb}{LR_c}\frac{1}{\sqrt{2\pi}}\text{exp}\left(-\frac{(a-bR_c)^2}{2}\right) \nonumber \\
&\quad -\frac{R_s}{2LR_c^2}\text{exp}\left(-\frac{(a-bR_c)^2}{2}\right),\\
\label{lemma22}		&=\frac{R_s}{LR_c^2}\text{exp}\left(-\frac{(a-bR_c)^2}{2}\right) \left(\frac{b}{\sqrt{2\pi}}R_c-\frac{1}{2} \right),
\end{align}
\end{small}
where inequality \eqref{lemma21} comes from the Chernoff bound of the Q-function, i.e., $Q(d)\leq \frac{1}{2}\text{exp}(-\frac{d^2}{2}),$ $ d\geq 0$, and $a-bR_c \geq0$.  From \eqref{lemma22}, we observe that when $R_c\geq \frac{\sqrt{2\pi}}{2b}$, the gradient $\frac{\partial U_{r}}{\partial R_c}\geq 0$. 
When $L$ tends to infinity, we have
\begin{align} \label{lemma23}
	\lim_{L \rightarrow \infty } \frac{\sqrt{2\pi}}{2b}\leq \lim_{L \rightarrow \infty }\frac{\sqrt{2\pi}\log_2(e)}{2\sqrt{L}}=0.
\end{align}
From \eqref{lemma23},  it is observed that $\frac{\sqrt{2\pi}}{2b}$ approximates to zero for large block length $L$. For example, when $L\geq 512$,  $\frac{\sqrt{2\pi}}{2b}\leq 0.07$.  
Hence, for sufficiently large $L$, we have $\frac{\partial U_{r}}{\partial R_c}\geq 0$, which implies that $U_r$ increases w.r.t. the channel rate $R_c$. Then, according to the constraint in problem \eqref{modelselec:23}, the optimal solution $R_c^*=\frac{R_s}{\tilde{d}}$. 
\addtolength{\topmargin}{0.01in}	

For the polar coding with the block error probability given in \eqref{polarc}, the objective function in problem \eqref{modelselec:23} becomes 
\begin{align}
		U_{p}=\frac{R_s}{LR_c}e^{\beta_1R_c+\beta_2}.
\end{align}
The gradient is calculated by
\begin{align}
	\frac{\partial U_p}{\partial R_c}&=\frac{R_s \beta_1}{LR_c}e^{\beta_1R_c+\beta_2}-\frac{R_s}{LR_c^2}e^{\beta_1R_c+\beta_2},\\
	\label{lemma24} &=\frac{R_s}{R_c^2L}e^{\beta_1R_c+\beta_2}\left(\beta_1 R_c-1 \right).
\end{align}
From \eqref{lemma24}, we observe that when $R_c\geq \frac{1}{\beta_1}$, $\frac{\partial U_p}{\partial R_c}\geq 0$. According to the constraint in problem \eqref{modelselec:23}, the optimal solution $R_c^*$ given in \eqref{lemma2:fuc1} is obtained. 

\section{Proof of Theorem \ref{Theorem1:1}}

When the NNs are given and the channel rate $R_c=\frac{R_s}{\tilde{d}}$,  the distortion $\mathcal{D}_c$ with the block error probability $\rho$ given in \eqref{randomc} can be expressed by
\begin{small}
\begin{align}
	&\mathcal{D}_c  = \frac{K}{M}\left( 1-(1-\rho)^{\tilde{T}}\right)\tilde{C}_{\psi} \left(\frac{2^{2R_s/K}}{2\pi e}+\frac{1}{12}\right),\\
\label{theoremp:1}	& = \frac{K}{M} \frac{\tilde{d}}{L} \tilde{C}_{\psi} \left(\frac{2^{2R_s/K}}{2\pi e}+\frac{1}{12}\right)Q\left(\frac{\sqrt{L}\left(\log_2(1+\gamma)-\frac{R_s}{\tilde{d}}\right)}{\sqrt{\left(1-\frac{1}{(1+\gamma)^2}\right)\log_2^2(e)}}\right), \\
\label{theoremp:2}	& \leq \frac{K}{M} \frac{\tilde{d}}{L}  \tilde{C}_{\psi} \left(\frac{2^{2R_s/K}}{2\pi e}+\frac{1}{12}\right)\frac{1}{2}\text{exp}\left(-\frac{L(\log_2(1+\gamma)-\frac{R_s}{\tilde{d}})^2}{2\log_2^2(e)}\right), 
\end{align}
\end{small}where \eqref{theoremp:1} comes from $(1-\rho)^{\tilde{T}} \approx 1-\tilde{T}\rho$ for small $\rho$ and $R_c=\frac{R_s}{\tilde{d}}$. Inequality \eqref{theoremp:2} comes from the Chernoff bound of the Q-function \cite{rohatgi2015introduction}, i.e., $Q(d)\leq \frac{1}{2}\text{exp}(-\frac{d^2}{2}),$ $ d\geq 0$,  $\log_2(1+\gamma)-\frac{R_s}{\tilde{d}}>0$, and $\left(1-\frac{1}{(1+\gamma)^2}\right)\leq 1$.  Since the NNs are fixed, the parameters $R_s$ and $ \tilde{C}_{\psi}$  are constant. 
Hence, there exist a positive constant $\eta$ with $\log_2(1+\gamma)-\frac{R_s}{\tilde{d}}>\eta$ such that the inequality \eqref{theoremp:2} holds. 
Then, Theorem \ref{Theorem1:1} is proved. 
\addtolength{\topmargin}{0.02in}	
	\IEEEpeerreviewmaketitle
\bibliographystyle{IEEEtran}
\bibliography{semantic}
\end{document}